\documentclass[reprint,
superscriptaddress,
 amsmath,amssymb,
 aps,
 prr,
]{revtex4-2}

\usepackage{graphicx}
\usepackage{dcolumn}
\usepackage{bm}

\usepackage[utf8]{inputenc}
\usepackage[T1]{fontenc}
\usepackage{mathtools}
\usepackage{etoolbox}
\usepackage{xcolor}
\usepackage{multirow}
\usepackage{bbm}
\usepackage[normalem]{ulem}
\usepackage{units}
\usepackage{placeins}
\usepackage{algorithm}
\usepackage{algpseudocode}
\usepackage{float}

\makeatletter
\def\@email#1#2{%
 \endgroup
 \patchcmd{\titleblock@produce}
  {\frontmatter@RRAPformat}
  {\frontmatter@RRAPformat{\produce@RRAP{*#1\href{mailto:#2}{#2}}}\frontmatter@RRAPformat}
  {}{}
}%
\makeatother

\usepackage{tikz}  

\newcommand{\koppatwo}{
\mathrel{%
    \vcenter{\offinterlineskip
      \ialign{##\cr\bigcirc\cr\noalign{\kern-1.5pt}|\cr}%
    }%
  }%
}

\usepackage{pdftexcmds}

\DeclareFontFamily{U}{cbgreek}{}
\DeclareFontShape{U}{cbgreek}{m}{n}{
        <-6>    grmn0500
        <6-7>   grmn0600
        <7-8>   grmn0700
        <8-9>   grmn0800
        <9-10>  grmn0900
        <10-12> grmn1000
        <12-17> grmn1200
        <17->   grmn1728
      }{}
\DeclareFontShape{U}{cbgreek}{bx}{n}{
        <-6>    grxn0500
        <6-7>   grxn0600
        <7-8>   grxn0700
        <8-9>   grxn0800
        <9-10>  grxn0900
        <10-12> grxn1000
        <12-17> grxn1200
        <17->   grxn1728
      }{}

\DeclareRobustCommand{\qoppa}{%
  \text{\usefont{U}{cbgreek}{\normalorbold}{n}\symbol{19}}%
}
\DeclareRobustCommand{\Qoppa}{%
  \text{\usefont{U}{cbgreek}{\normalorbold}{n}\symbol{21}}%
}
\makeatletter
\newcommand{\normalorbold}{%
  \ifnum\pdf@strcmp{\math@version}{bold}=\z@ bx\else m\fi
}
\makeatother

\newcommand {\com}[1]{{\color{black} #1}}

\usepackage{amsthm}

\begin{document}


\title[Circulance]{\com{Quantifying the irregularity of a time series}} 

\author{Max Potratzki}
\affiliation{Department of Epileptology, University of Bonn Medical Centre, Venusberg Campus 1, 53127 Bonn, Germany}
\affiliation{Helmholtz Institute for Radiation and Nuclear Physics, University of Bonn, Nussallee 14--16, 53115 Bonn, Germany}

\author{Manuel Adams}
\affiliation{Department of Epileptology, University of Bonn Medical Centre, Venusberg Campus 1, 53127 Bonn, Germany}
\affiliation{Helmholtz Institute for Radiation and Nuclear Physics, University of Bonn, Nussallee 14--16, 53115 Bonn, Germany}

\author{Timo Br\"ohl}
\affiliation{Department of Epileptology, University of Bonn Medical Centre, Venusberg Campus 1, 53127 Bonn, Germany}
\affiliation{Helmholtz Institute for Radiation and Nuclear Physics, University of Bonn, Nussallee 14--16, 53115 Bonn, Germany}

\author{Klaus Lehnertz}
\email{klaus.lehnertz@ukbonn.de}
\affiliation{Department of Epileptology, University of Bonn Medical Centre, Venusberg Campus 1, 53127 Bonn, Germany}
\affiliation{Helmholtz Institute for Radiation and Nuclear Physics, University of Bonn, Nussallee 14--16, 53115 Bonn, Germany}
\affiliation{Interdisciplinary Center for Complex Systems, University of Bonn, Br{\"u}hler Stra\ss{}e 7, 53175 Bonn, Germany}

\date{\today}

\begin{abstract}
We introduce circulance, a scalar measure for classifying time series of dynamical systems.
Circulance captures the extent of temporal regularity or irregularity that is encoded in the topology of a directed ordinal pattern transition network derived from a time series.
We demonstrate numerically that circulance sensitively and robustly positions time series \com{of canonical model systems, representative of preset dynamical regimes,} along a continuous spectrum from regularity to randomness.
\com{Analyzing empirical data from long-term observations of high-dimensional, complex systems -- human brain and the Sun -- reveals that circulance aids in elucidating different dynamical regimes.}
\end{abstract} 

\pacs{}

\maketitle 

\section{Introduction}
Investigations of dynamical systems often depend on time series data.
A central task is to classify the behavior of a system and how it may evolve over time. 
\com{Identifying} regularity \com{in} a time series and its potential breakdown is crucial for understanding the underlying dynamics, distinguishing between periodic, quasiperiodic, chaotic, and stochastic behavior, and for assessing a system's stability.
However, identifying regularity in practice is fraught with difficulties.
Empirical time series may be contaminated with measurement noise and other unwanted external \com{disturbances} and may reflect transient dynamics. 
These challenges are compounded when deviations from regularity are subtle, gradual, or localized in time. 
Beyond classifying a system's dynamics as regular or not, it is therefore of high interest to position the dynamics along a spectrum ranging from strictly periodic via chaotic to fully stochastic.

\com{Several methods for the classification of time series have been proposed and critically discussed}~\cite{kantz2003,sun2007,lacasa2008,lacasa2010,Bloomfield2004,gottwald2004,marwan2007,freitas2009,kulp2011,kulp2014,mateos2017,Zou2019,tabar2019book,tempelman2020,Hamilton2020,kottlarz2023,angelidis2024}.  
Although powerful in many contexts, these methods typically focus on detecting the presence of dominant periodic components in either the time domain, frequency domain, the system's reconstructed phase space or by representing the time series as a network. 
Only a few methods provide a classification \com{and only} within parts of the continuum from strictly periodic to fully stochastic~\cite{small2001,luo2005,zou2007,Rosso2007,gao2009,luque2009,marwan2009,zou2010,ngamga2012,das2016,olivares2020,zhang2021,boaretto2021,prado2022,ren2023,fouda2024,mohan2025}\com{, yet not the full spectrum}.
Moreover only few methods are explicitly designed to handle data from nonstationary systems~\cite{priestley1988,manuca1996,fokianos1998,hegger2000,rieke2002,fryzlewicz2009,lepage2009}.
These limitations necessitate the development of a flexible and advanced approach to time series classification.

\com{Here, we present a novel approach for a quantitative classification of a time series derived from a suitable observable of a complex dynamical system.}
\com{It has been shown that probabilities of transitions between ordinal patterns derived from time series encapsulate vital information about a system's dynamics~\cite{bandt2002,unakafov2014,Zhang2017a,cuesta2019,zanin2021}.
These transitions can be topologically represented as ordinal pattern transition networks (OPTNs)~\cite{small2013,mccullough2015,olivares2020,zanin2021}.}
We leverage graph-theoretical metrics -- fundamentally a vertex' in- and out-degree -- to investigate the intricate topological properties of these networks constructed from time series data.
Specifically, we assess to what extent an OPTN's topology \com{deviates from a circular structure} and quantify this extent with \com{a single scalar measure, to which we refer as circulance in the following.} 
We demonstrate that an OPTN's circulance encodes essential dynamical aspects of a time series, which allows \com{a quantitative distinguishing of} dynamical regimes across diverse synthetic and \com{empirical} datasets.

\section{Methods}
\subsection{From a Time Series to an OPTN}
\begin{figure*}[htbp]
    \centering
    \includegraphics[width=\linewidth]{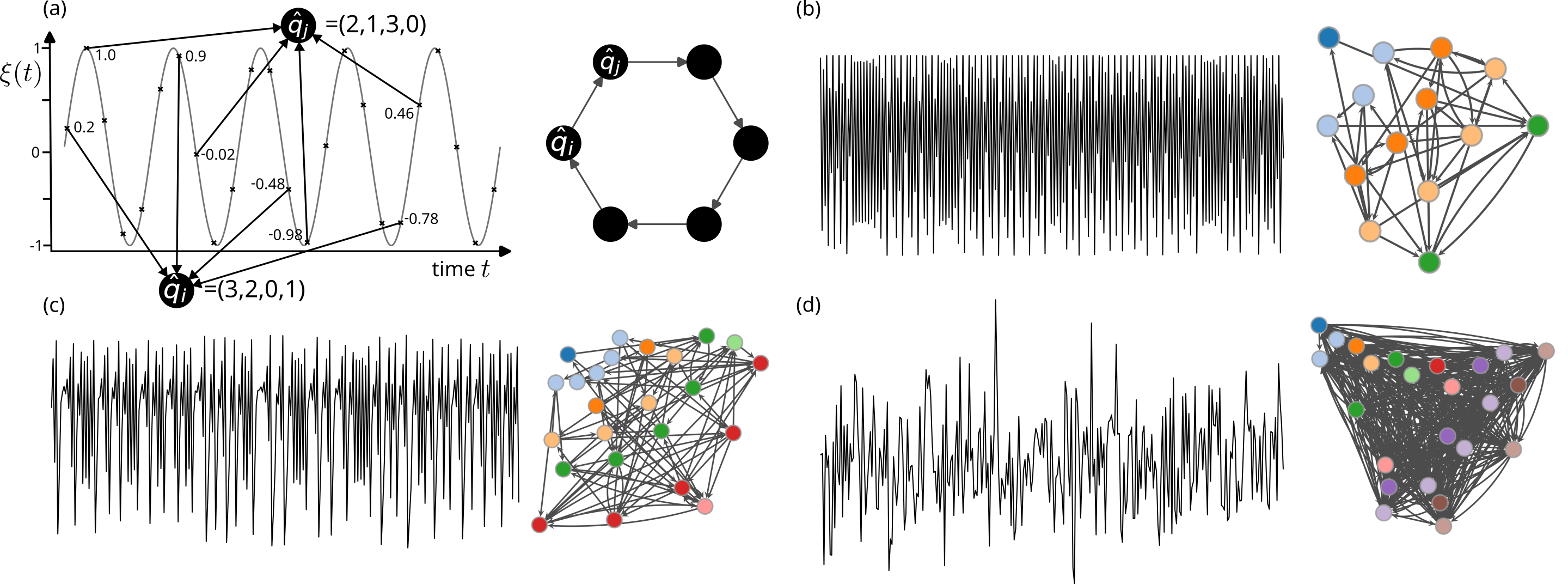}
    \caption{\com{Different types of dynamics and the corresponding ordinal pattern transition networks (OPTNs).
   	Within each subfigure, vertices of the same color have the same number of adjacent vertices.
   	(a) Schematic of the construction of an OPTN from a strictly periodic dynamics (with $d=4$ and $\tau=1$). The resulting OPTN consists of one closed trail or circuit on the OPTN. (b) A quasiperiodic dynamics results in a broader degree distribution (more distinct colors) of the OPTN. (c) Chaotic and (d) stochastic dynamics result in even more vertices and a higher amount of vertices with a distinct number of adjacent vertices in the respective OPTN. For a stochastic dynamics, the deviation from a simple circuit is maximum.
   	Time series shown in (b)-(d) are of same length $N$.}}
    \label{fig:scheme2}
\end{figure*}

\com{A time series $\xi(t)$, with $t\in\mathbb{N}_0^+$, $t\in[0,N-1]$, is divided into partitions $p(s)$ (with running index $s\in\{0,1,\dots,N-1-(d-1)\tau\}$) of size $d$ (embedding dimension), with each of the elements of $p(s)$ being separated by $\tau$ time steps (embedding delay).
From each partition $p(s)|_{s=t'}$, an ordinal pattern $q(s)|_{s=t'}$ is inferred by a rank-ordering of the elements of the partition~\cite{bandt2002} $p(s)|_{s=\{t'\}}=\left[\xi(t'),\xi(t'+\tau),\dots,\xi(t'+(d-1)\tau)\right]$, where $t'\in\{0,1,\dots,N-1-(d-1)\tau\}$.
With $\mathcal{X}_{t'}=\{\xi(t'),\xi(t'+\tau),\dots,\xi(t'+(d-1)\tau)\}$, the rank ordering function $\rho^*$ can be defined as
\begin{equation*}
	\rho^*: \mathcal{X}_{t'} \rightarrow \mathbb{N}_0,
\end{equation*}
such that
\begin{equation}
	\rho^*(u) = |\{v\in \mathcal{X}_{t'}|u<v\}|,
\end{equation}
where $u,v\in\mathcal{X}_{t'}$.
Therefore, $q(s)$ is a sequence of unique ordinal patterns.
 As $d!$ unique ordinal patterns are possible, the unique ordinal pattern $\hat{q}_m$ ($m\in[0,d!-1]$) is obtained by ordering the unity ordinal pattern $\hat{q}_\mathbbm{1} = (0,1,\dots,d-1)$ according to the permutation $\rho(\mathcal{X}_{t'})$:
\begin{equation*}
	\rho: \mathcal{X}_{t'}^d\rightarrow\mathbb{N}_0^d,
\end{equation*}
with
\begin{equation}
	\begin{aligned}
		\rho(\mathcal{X}_{t'})=(\rho^*(\mathcal{\xi}(t')),\rho^*(\xi(t'+\tau)),\\\ldots,\rho^*(\xi(t'+(d-1)\tau))),
	\end{aligned}
\end{equation}
such that
\begin{equation}
	\begin{aligned}
		\hat{q}_i &=(0,1,\dots,d-1)_{\rho(\mathcal{X}_{t'})}\\
		&=(0_{\rho^*(\xi(t'))},1_{\rho^*(\xi(t'+\tau))},\dots,(d-1)_{\rho^*(\xi(t'+(d-1)\tau))}).
	\end{aligned}
\end{equation}

To illustrate the derivation of a unique ordinal pattern, consider the partition $p(s) = (0.2,0.9,-0.48,-0.78)$, whose elements form the set $\mathcal{X}_{t'}~=~\{0.2,0.9,-0.48,-0.78\}$, for some $s$ and some $t'$  (cf. Fig.~\ref{fig:scheme2} (a)).
Now as $\rho(\mathcal{X}) = (2,3,1,0)$, this results in the corresponding  unique ordinal pattern $\hat{q}_m~=~(0_2,1_3,2_1,3_0) = (3,2,0,1)$ for some $m$.
By convention, in the case of repeated amplitude values ranks are assigned according to the order of temporal occurrence, e.g., $p(s) = (1,1,1,1)$ leads to $\hat{q}_m = \hat{q}_\mathbbm{1}$.
Therefore, constant amplitudes yield the same unique ordinal pattern as strictly increasing amplitudes.\\
}

\com{An OPTN can be represented by a directed graph $\mathcal{O}=\{\mathcal{V},\mathcal{E}\}$ consisting of a set of vertices $\mathcal{V}$ (with $V=|\mathcal{V}|$ elements) and a set of directed edges $\mathcal{E}$ (with $E=|\mathcal{E}|$ elements).
Having transformed a time series to a sequence of ordinal patterns, we can now construct a topological representation of the transitions between successive patterns~\cite{mccullough2015}.
To do so, each unique ordinal pattern $\hat{q}_i$ is associated with a vertex $v_i\in\mathcal{V}$ ($i\in[0,V-1]$) in the OPTN. 
An edge $e_{ij}\in\mathcal{E}$ exists between vertices $v_i$ and $v_j$, if there is a transition from pattern $\hat{q}_i$ to pattern $\hat{q}_j$ in the temporal sequence of ordinal patterns $q(s)$.
Therefore, vital topological aspects of the OPTN can be represented \com{as} an adjacency matrix $\mathbf{A}$ with $A_{ij}= 1$ if $e_{ij}$ exists, and $A_{ij} = 0$ otherwise.
Self-loops --~transitions from a pattern $\hat{q}_i$ onto itself~-- are not considered ($A_{ii}=0$), as they do not represent a change in the prevailing dynamics.
The (ir)regularity of a time series is translated to a topological level of an OPTN, as shown in Figure~\ref{fig:scheme2}.   
}

\subsection{\com{Defining Circulance}}
\subsubsection{\com{Capturing distinct transitions between unique ordinal patterns}}
The in- and \com{the} out-degree $k^\mp_{v_i}$ of vertex $v_i$ in an OPTN assess the amount of unique ordinal patterns directly preceding (in-degree) respectively following (out-degree) \com{the unique ordinal pattern $\hat{q}_i$ associated with $v_i$ and are defined as }
\begin{align}
    k^-_{v_i} = \sum_{v_j \in \mathcal{V}} A_{ji}, \;\;\;k^+_{v_i} = \sum_{v_j \in \mathcal{V}} A_{ij}.
\end{align}
\com{Since an OPTN retains essential aspects of a time series~\cite{small2013} -- provided embedding parameters $d$ and $\tau$ were chosen appropriately --} we conjecture that the in- and out-degrees provide a means of characterizing this time series and potentially differentiating between dynamical regimes.
Generally, each dynamics exhibited by the system over a given time interval exceeding $(d-1)\tau$ may be represented as a sequence of ordinal patterns and therefore as a walk on the OPTN.
Strictly periodic dynamics translate to a chordless cycle (Fig.~\ref{fig:scheme2}).

The possibility to pursue a closed trail in a directed network is largely reflected by the vertices' in- and out-degrees.
We therefore consider the \com{normalized} in- and out-degree distributions $\Gamma^\mp(k^\mp)$ of the OPTN to exhibit specific properties depending on the system's dynamics.
A strictly periodic time series satisfies the condition $\xi(t+T) = \xi(T)$ with period $T\in\mathbb{N}^+$, translating to a repeating temporal sequence of ordinal patterns.
This resembles a walk in circles on the OPTN, as further reflected by the OPTN's circular structure with all vertices having the same in- and out-degrees $k=k^+=k^-$ (excluding the case of a fully connected network). 
Consequently, the normalized degree distributions $\Gamma^\mp(k^\mp)$ take the form $\Gamma^\mp(k^\mp) = \delta_k(\mathcal{K}^\pm)$,  
where $\mathcal{K}^\pm$ denote the sets of in-/out-degrees, with $k^\pm\in\mathcal{K^\pm}$ and $k\in\mathbb{N}^+$, and $ \delta_k $ denotes the Dirac measure, such that
\begin{align}
\sum_{v_i\in\mathcal{V}}k^-_{v_i} = \sum_{v_i\in\mathcal{V}}k^+_{v_i} = |\mathcal{E}|.
\end{align}
Hence, for a strictly periodic time series -- representing the left end of the spectrum of dynamical regimes -- it then holds $ |\mathcal{K}^+|=|\mathcal{K}^-|=1$ .

Deviations from strict periodicity lead to an increase in the number of different in- and out-degrees, depending on the specific dynamics \com{(encoded as different colors in Fig.~\ref{fig:scheme2})}.  
\com{We refer to the amount of degree values that are contained in $\mathcal{K^+}$ while also being in $\mathcal{K^-}$ as (qoppa $\qoppa$):}
\begin{align}
 \qoppa_\tau=|\mathcal{K}^+\cap\mathcal{K}^-|.
\end{align}
This property has an indirect dependence on the embedding delay $\tau$ as the latter can determine topological aspects of the OPTN such as the in- and out-degrees of the vertices.
We recall that we generally have to assume the time series sufficiently captured a suitable observable of the underlying system dynamics to allow any conclusion about the dynamical system itself.
This is not universally given, \com{particularly not for high-dimensional dynamical systems.}
\com{A purely stochastic \com{time series}, \com{such as that of a white noise process}, represents the other extreme (right end) in the spectrum of dynamical regimes, and therefore poses a reasonable reference~\cite{pessa2019} to determine a sufficient embedding dimension for the construction of an OPTN.}
\subsubsection{\com{Influence of time series length and embedding parameters}}
For a stochastic \com{time series} of length $N\sim N_q$, $N_q\rightarrow\infty$, an embedding dimension $d$ would translate to an OPTN with each of the $V=d!$ possible \com{unique} ordinal patterns constituting a vertex.
Consequently, it is to be expected that every possible transition between all pairs of \com{unique} ordinal patterns is observed in the temporal sequence of ordinal patterns, resulting in $E_{\rm max}=V(V-1)=d!(d!-1)\simeq d!^2$ edges in the OPTN.

Let us consider a case where $N_q\gg E_{\rm max}$.
Consequently, the in- and out-degree distributions of the respective OPTN are rather homogeneous, as the OPTN is densely connected and approaches a fully connected network.
Therefore, each vertex has approximately the same large degree.
An opposing case would be constituted by $N_q\ll E_{\rm max}$. 
Here, the number of both unique ordinal patterns and unique transitions present in the temporal sequence of ordinal patterns does not even approximately amount to the number of all possible \com{unique} ordinal patterns and transitions.
\com{Only a very small subset of all possible transitions can be observed, and each of these transitions occurs randomly and rarely, resulting in a sparse OPTN with few low-degree vertices and a homogeneous degree distribution.}

In between the two cases, we can further assume a scenario for which $N_q\sim E_{\rm max}$ holds.
Only a fraction of possible \com{unique} ordinal patterns and transitions is observed.
Different transitions are unevenly distributed, translating to a heterogeneous degree distribution of the respective OPTN.
Consequently, and to guarantee maximum separability between a chaotic and a stochastic \com{time series} of length $N$ we choose the embedding dimension $d$, such that 
\begin{equation}
    N\sim d!^2.
    \label{eq:upper_bound}
\end{equation}
\begin{figure*}[htbp]
	\centering
	\includegraphics[width=\linewidth]{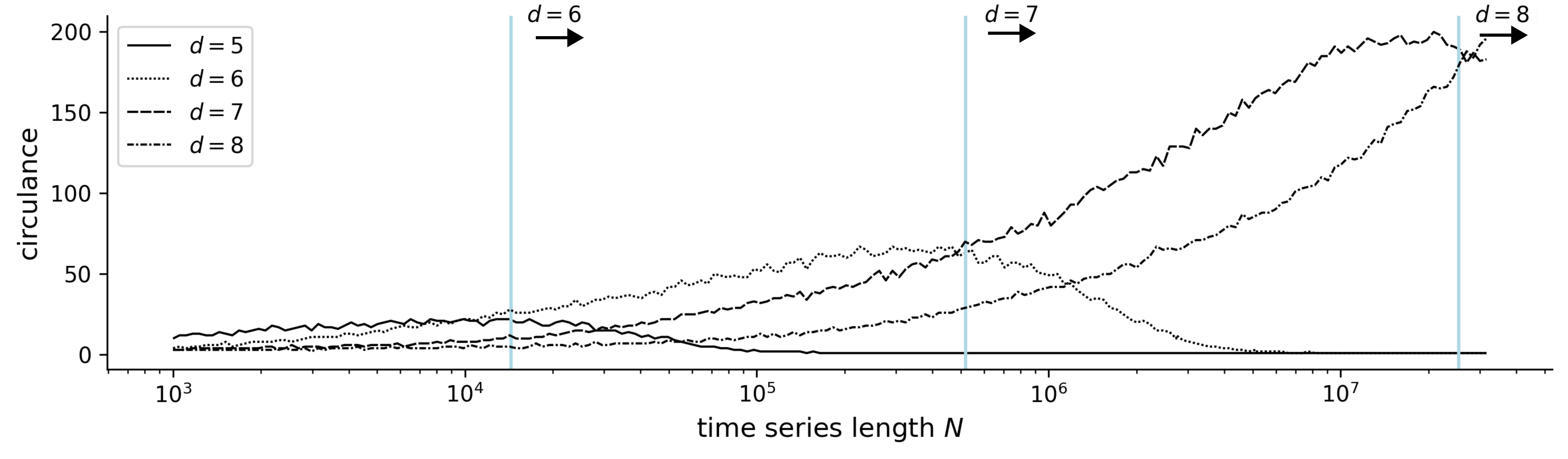}
	\caption{Circulance $\Qoppa$ in dependence of time series length $N$ for a single realization of a stochastic process and different embedding dimensions $d$.
	For a given time series length $N$, the embedding dimension $d$ is chosen such that circulance is maximum.
	Blue vertical lines mark $N=d!^2$ (cf. Eq.~\ref{eq:upper_bound}).
	}
	\label{fig:stoch_d}
\end{figure*}

Due to the ambiguity of choosing an embedding delay,
we propose to calculate $\qoppa_\tau$ for OPTNs constructed with
embedding delays $\tau \in \left[2,\ldots,\tau_{\rm{max}}\right]$, with $\tau_{\mathrm{max}}$ chosen such that at most $10$\% of the number of data points is lost (i.e., $(d-1)\tau_{\mathrm{max}}=N/10$) \com{when deriving the ordinal pattern sequence.}
\subsubsection{\com{Circulance}}
\com{For the final definition of circulance, we consider the time delay that minimizes $\qoppa_\tau$, since periodicity can not be artificially induced by an inappropriate choice of the embedding delay~\cite{kantz2003}. Eventually, with a sufficiently chosen embedding dimension $d$, we define circulance $\Qoppa\in\mathbb{N}^+$ as (Qoppa $\Qoppa$):}
\begin{equation}
    \Qoppa=\min{\qoppa_\tau}.
\end{equation}
The less regular (or periodic) a \com{time series} is the larger is the \com{time series}' circulance.
The left end of the spectrum of dynamical regimes, that range from periodic over quasiperiodic and chaotic to stochastic, is located at $\Qoppa=1$, independent of the length of the investigated \com{time series}, provided that at least one period has been captured.
The right end of the spectrum does depend on the length of the investigated \com{time series}, and is determined by $\Qoppa$ of a stochastic \com{time series} of the same length (cf. Eq.~\ref{eq:upper_bound} and Fig.~\ref{fig:stoch_d}).

\begin{table}[htbp]
    \centering
    \begin{tabular}{c|c|c|c|c|c}
            dynamics/system& $\tau_{\mathrm{min}}$&$V$    &$E$    &$\Qoppa$& observable \\
        \hline
         \textbf{\vspace{-0.35cm}superimposed}&$2$&$18$&$18$&$1$&\raisebox{-.7\height}{\includegraphics[width=2cm]{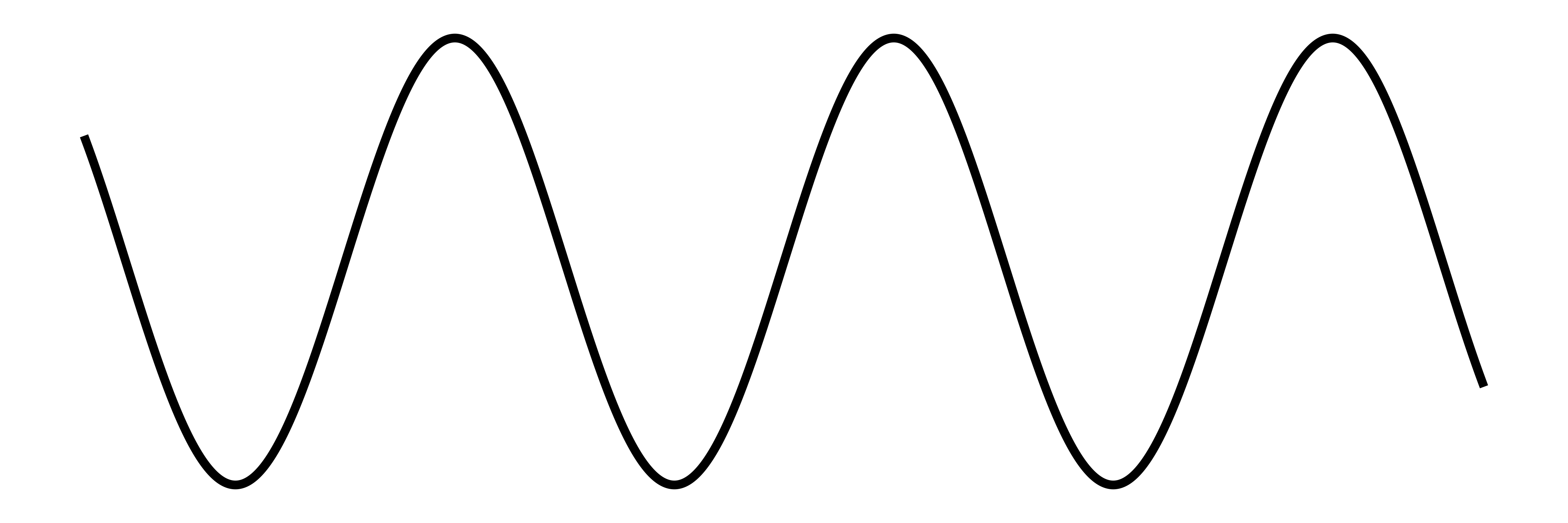}}  \\ 
         \textbf{sine waves}&&&&&\\
         &$2$&$18$&$18$&$1$&\raisebox{-.3\height}{\includegraphics[width=2cm]{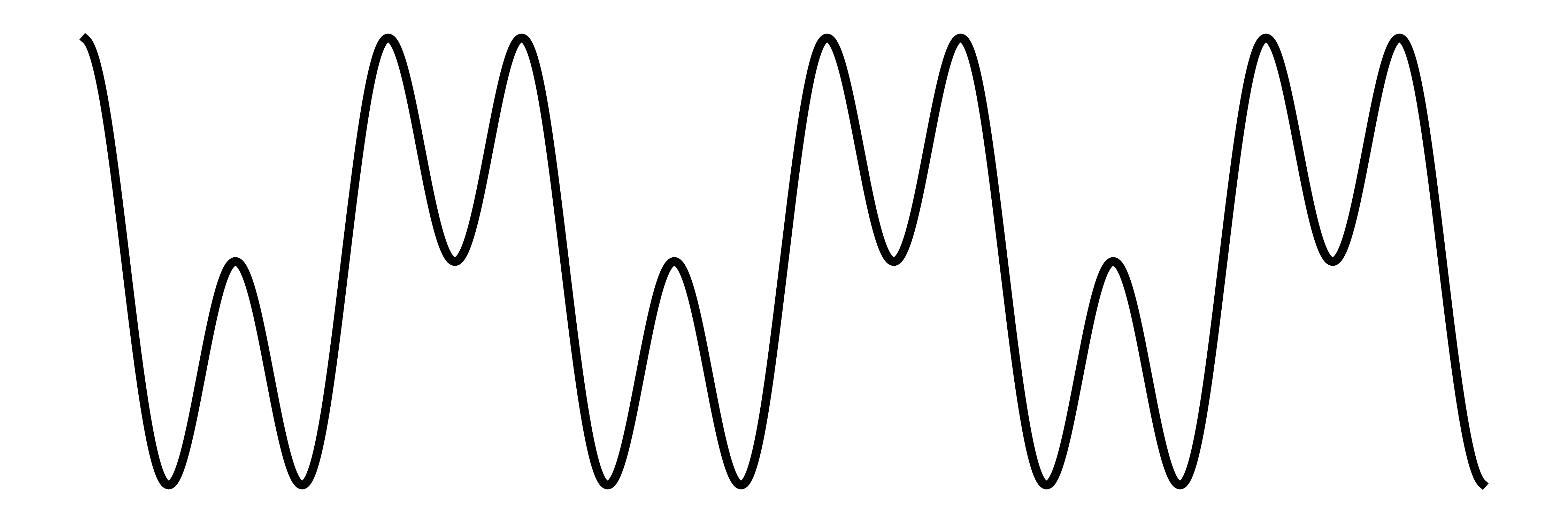}}  \\
         &$2$&$18$&$18$&$1$&\raisebox{-.35\height}{\includegraphics[width=2cm]{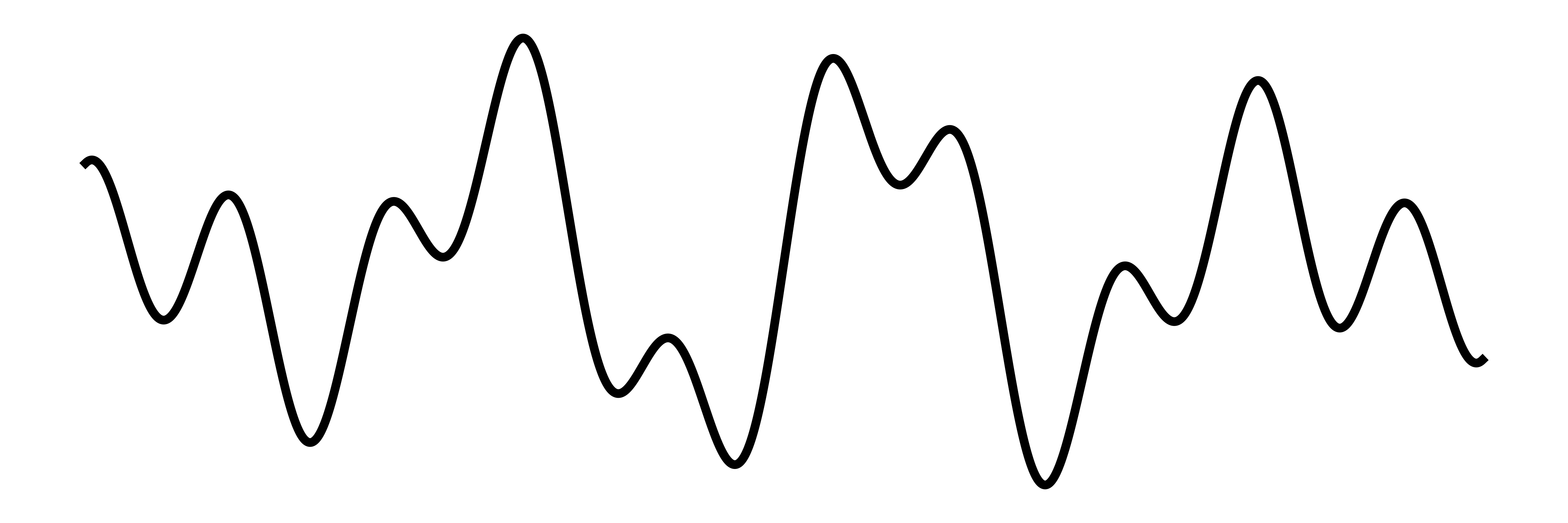}}  \\
         &&&&&\\
         
         \textbf{sawtooth}&2&18&18&1&\raisebox{-.35\height}{\includegraphics[width=2cm]{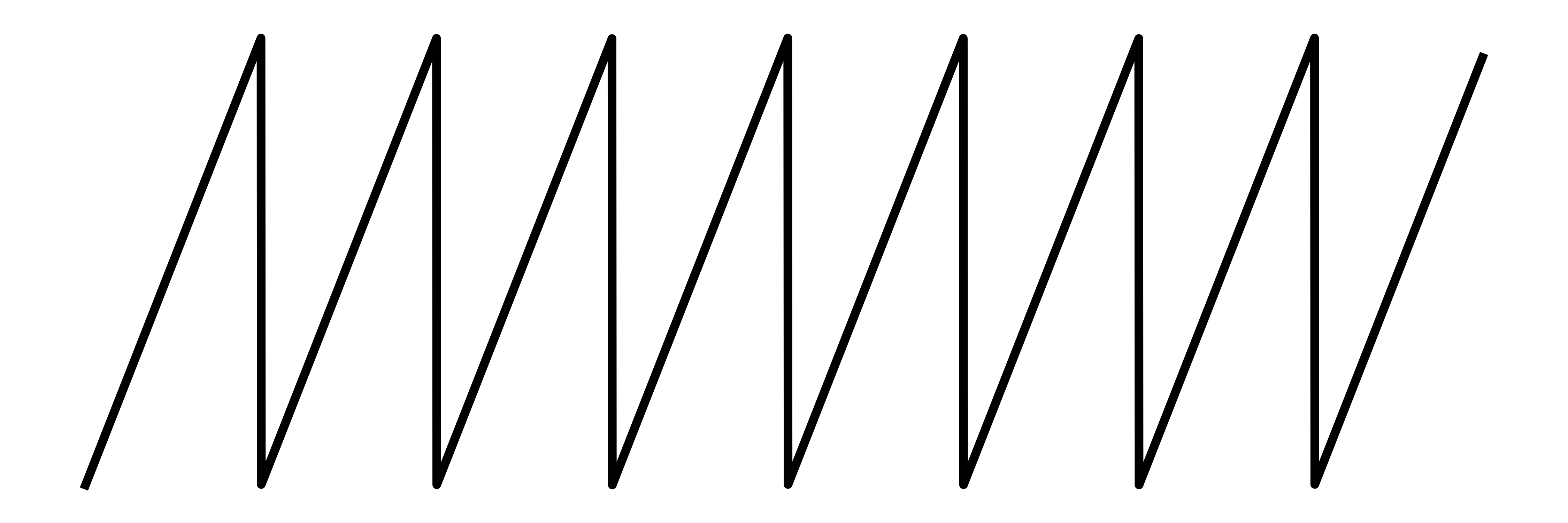}}  \\
         &&&&&\\

         \textbf{v. d. Pol oscillator}&&&&&$x$-component  \\
         periodic&7&30&30&1&\raisebox{-.35\height}{\includegraphics[width=2cm]{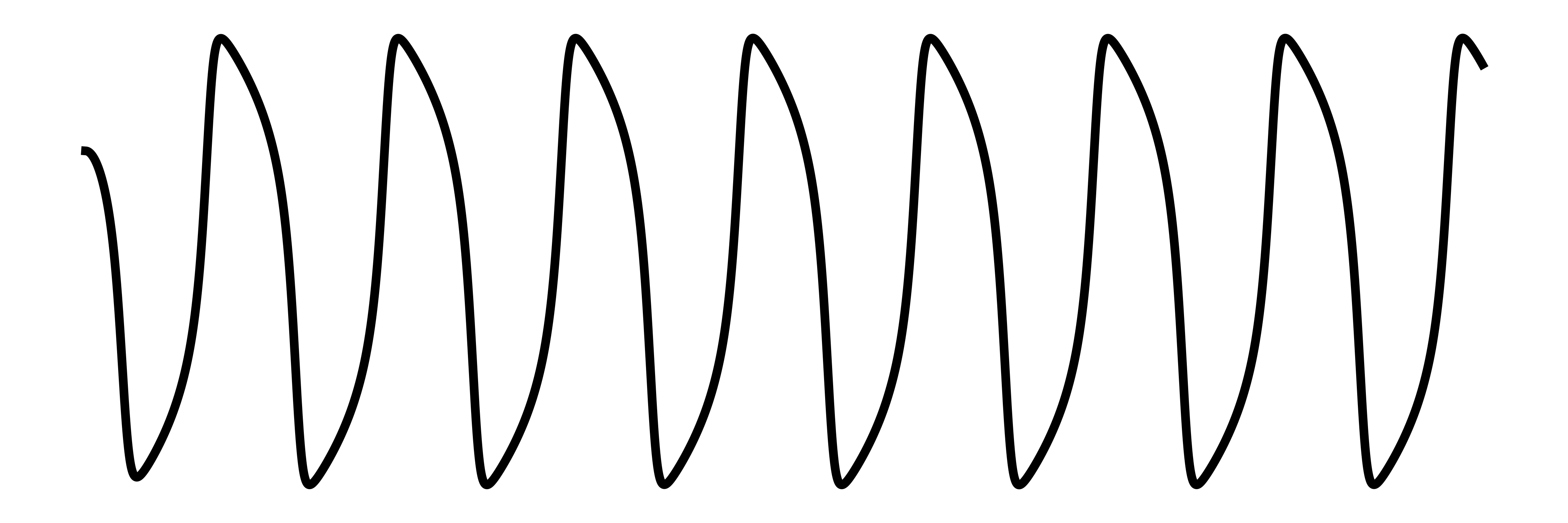}}\\
         periodic&12&30&30&1&\raisebox{-.35\height}{\includegraphics[width=2cm]{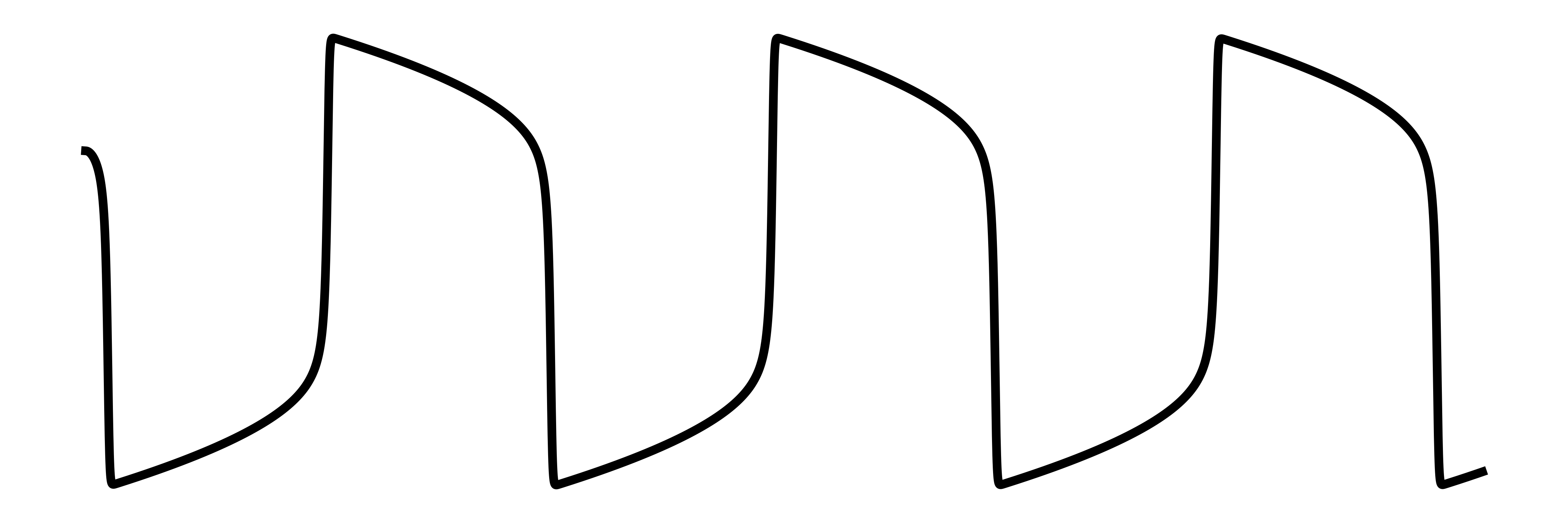}}\\
         &&&&&\\
         
         \textbf{FHN oscillator}&&&&&$x$-component  \\
         quasiperiodic&6&38&75&3&\raisebox{-.35\height}{\includegraphics[width=2cm]{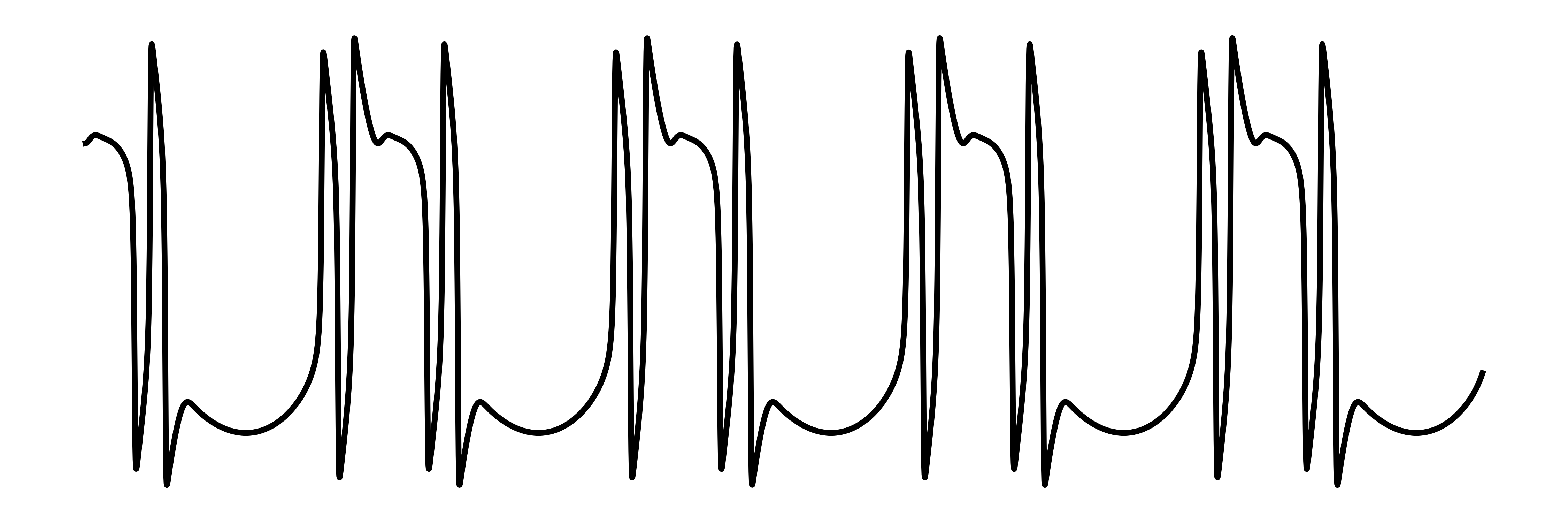}}\\
         quasiperiodic&183&84&151&3&\raisebox{-.35\height}{\includegraphics[width=2cm]{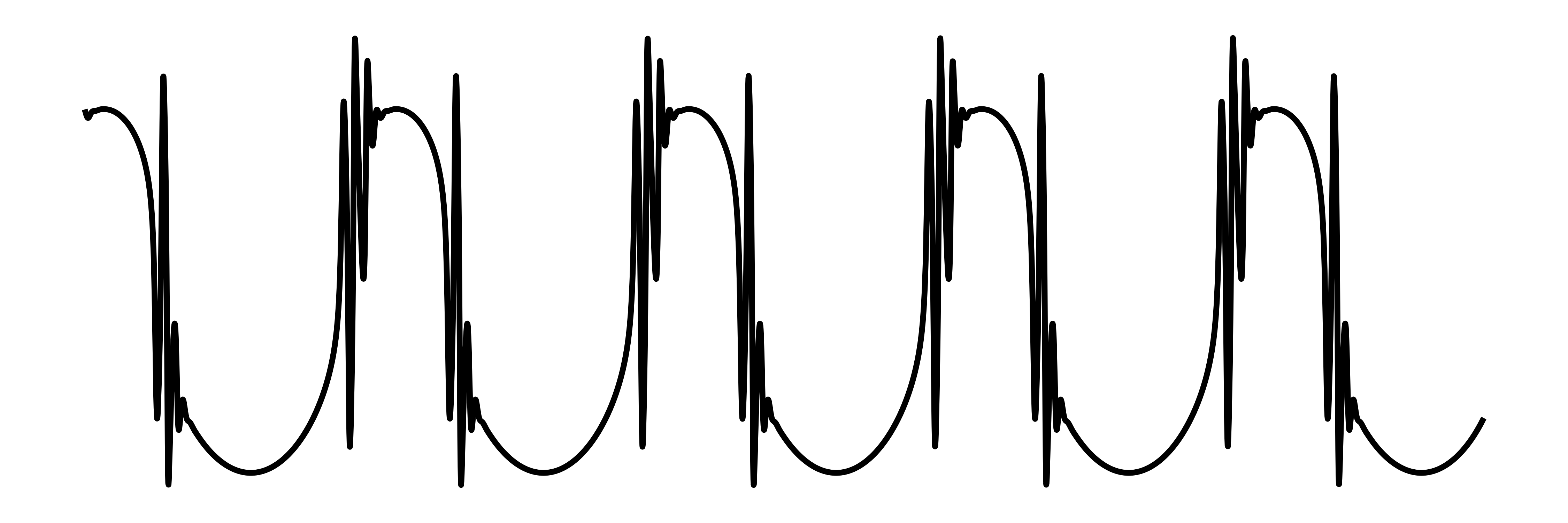}}\\
         &&&&&\\

         \textbf{coupled FHN osc.}&&&&&mean $x$-comp.\\
         quasiperiodic&17&30&43&2&\raisebox{-.35\height}{\includegraphics[width=2cm]{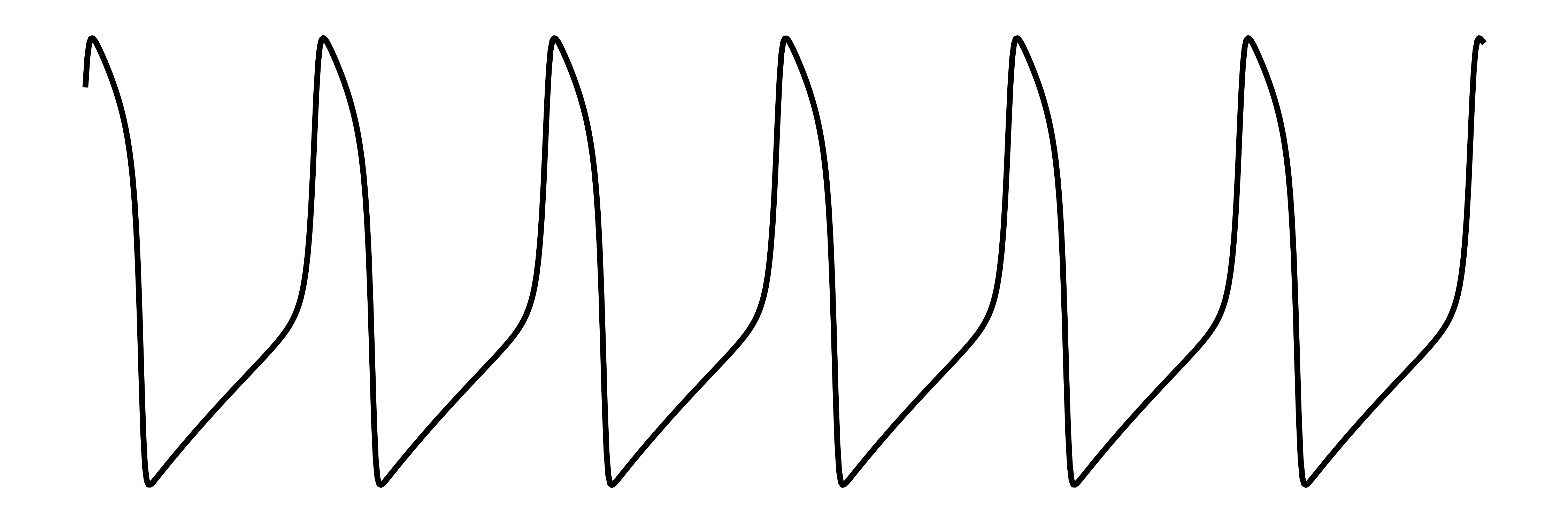}}\\
         quasiperiodic&9&268&456&4&\raisebox{-.35\height}{\includegraphics[width=2cm]{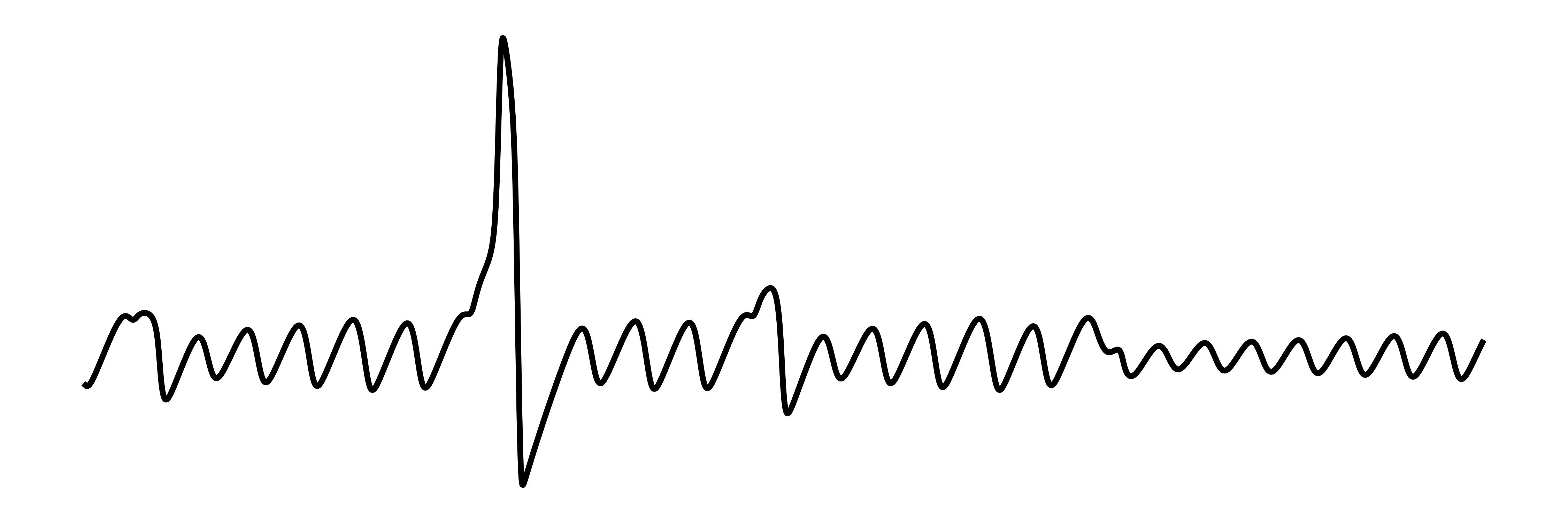}}\\
         chaotic&26&708&2787&8&\raisebox{-.35\height}{\includegraphics[width=2cm]{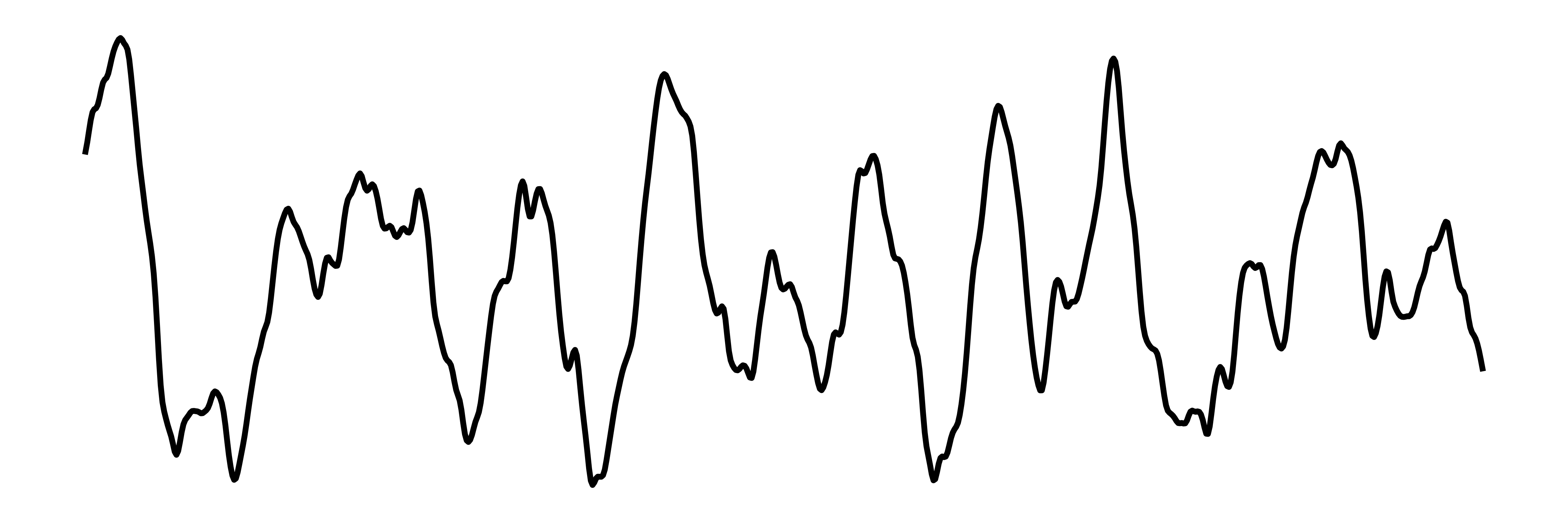}}\\
         &&&&&\\
         
         \textbf{Hénon map} &&&&&$x$-component\\
         quasiperiodic&3&32&76&4&\raisebox{-.35\height}{\includegraphics[width=2cm]{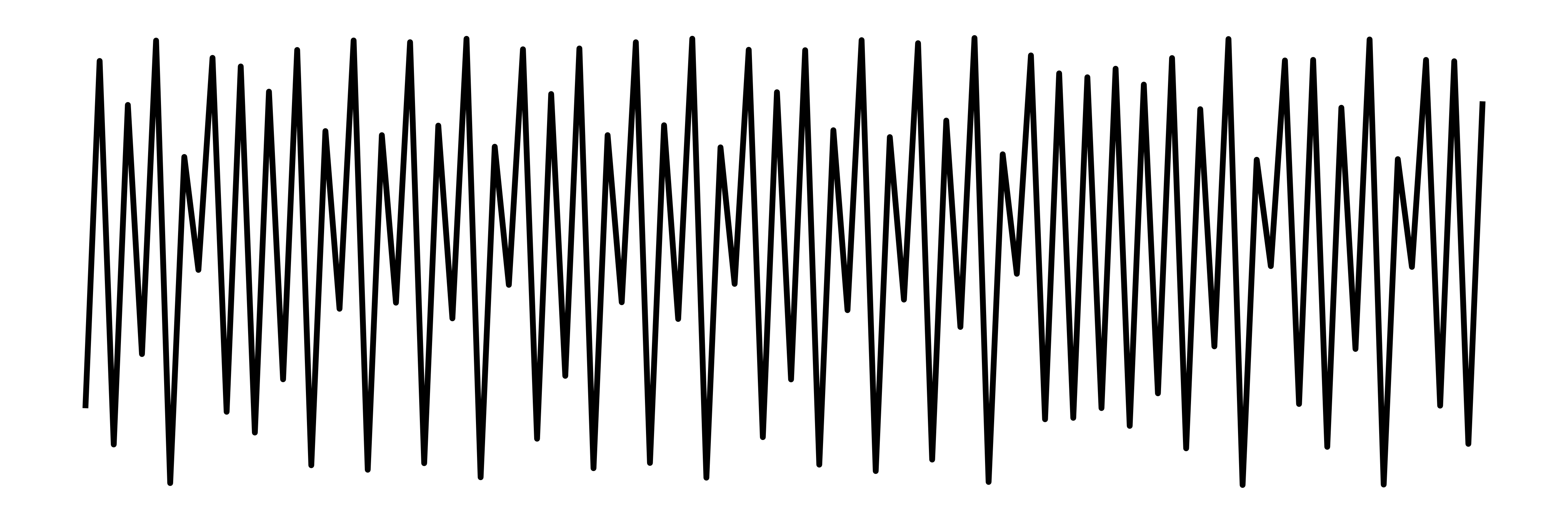}}\\
         chaotic&3&284&1142&\com{7}&\raisebox{-.35\height}{\includegraphics[width=2cm]{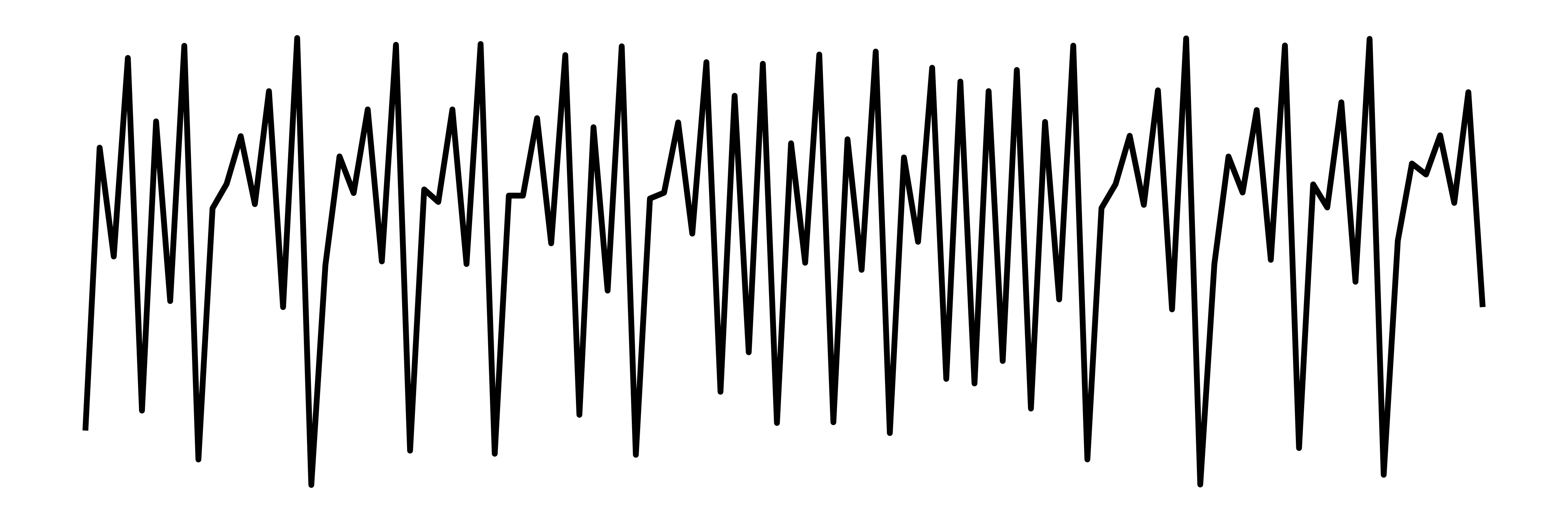}}\\
         chaotic&53&720&8429&\com{10}&\raisebox{-.35\height}{\includegraphics[width=2cm]{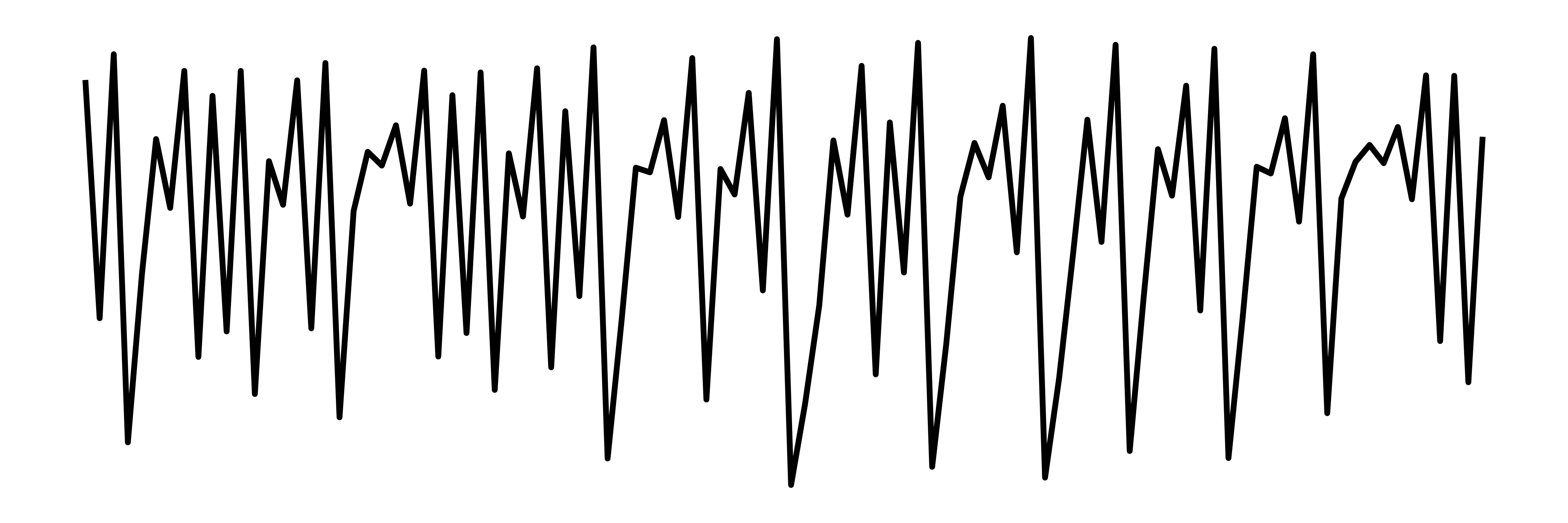}}\\
         &&&&&\\
         
         \textbf{Lorenz oscillator}&&&&&$x$-component\\
         quasiperiodic &13&66&80&2&\raisebox{-.35\height}{\includegraphics[width=2cm]{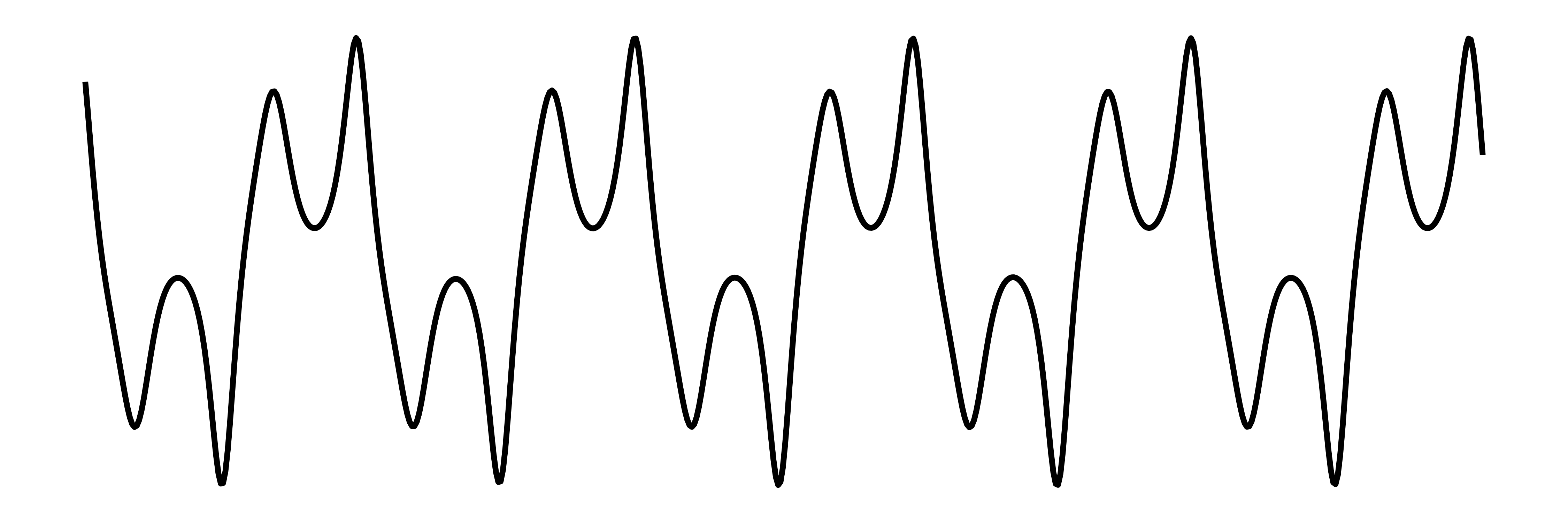}}\\ 
         quasiperiodic &2&42&81&4&\raisebox{-.35\height}{\includegraphics[width=2cm]{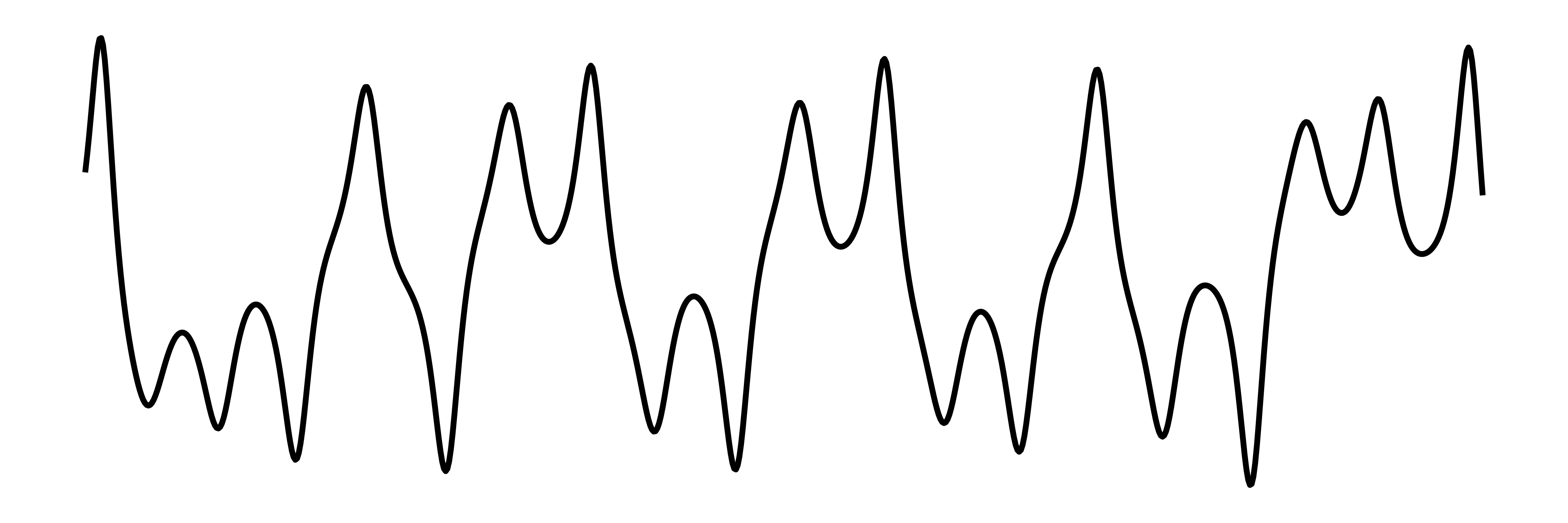}}\\ 
         &&&&&\\
         
         \textbf{stochastic}&16&720&9793&\com{18}&\raisebox{-.35\height}{\includegraphics[width=2cm]{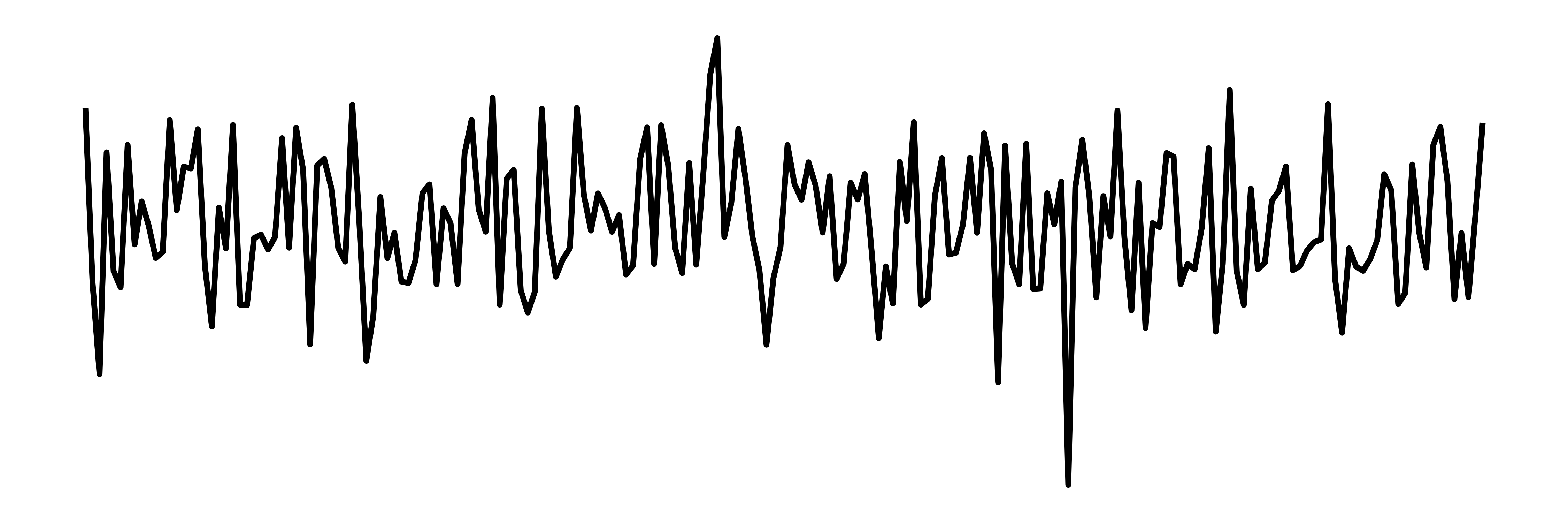}}\\
              
    \end{tabular}
    \caption{
    \com{Circulance $\Qoppa$ for different types of known dynamics of well-studied model systems for appropriately chosen control parameters (see Appendix~\ref{app:mod})}.
    Time series had a length $N = 10000$, which led us to choose an embedding dimension $d=6$.
    $\tau_{\rm min}$ denotes the embedding delay that minimizes $\qoppa_\tau$, and $V$ and $E$ are the number of vertices and edges of the respective OPTN.}
    \label{tab:dynamics_circulance}
\end{table}

\com{\section{Investigated Systems}
\subsection{Model Systems}
We validate our approach using time series generated from various canonical dynamical systems (see Appendix~\ref{app:mod}), capable of exhibiting distinct dynamical regimes \com{in the continuum} from strictly periodic over chaotic to stochastic\com{, predetermined by the choice of the systems' control parameters}.
Embedding dimensions are chosen based on the lengths of the respective time series (see Figure~\ref{fig:stoch_d}).
All continuous-time systems are integrated using a fourth-order Runge-Kutta method with adaptive step size control. 
Sampling rates are chosen to ensure adequate temporal resolution while avoiding oversampling artifacts.
For discrete maps, iterations are performed directly.
Time series have a length $N$ after discarding transients. 

\subsection{Real World Systems}
To assess the applicability of circulance to empirical data, we here analyze time series of two high-dimensional, complex systems -- human brain and the Sun -- for which long-lasting recordings of their dynamics are available.
The exact nature of these dynamics is matter of ongoing debate, and previous research often renders the respective dynamics seemingly stochastic.
We here investigate whether circulance can aid in elucidating the intricate characteristics of such complex dynamics.
\subsubsection{Human Brain Dynamics}
As a first example, we consider multichannel EEG data that was continuously recorded for three days from a healthy subject.
The subject had signed informed consent that their clinical data might be used and published for research purposes, and the study protocol had previously been approved by the local ethics committee.
EEG electrodes were placed according to the International 10-20 system~\cite{Seeck2017} with position Cz serving as physical reference.
Data were sampled at \unit[256]{Hz} using a 16 bit analog-to-digital converter and bandpass filtered offline between \unit[1~-~45]{Hz} (4th order Butterworth characteristic). 
A notch filter (3rd order) was used to suppress contributions at the line frequency (\unit[50]{Hz}). 
We visually inspected the continuous recording for strong artifacts (e.g., subject movements or amplifier saturation) and excluded such data from further analyses.
\subsubsection{Solar Magnetic Activity}
As a second example, we consider multi-decade sunspot number records, associated with solar magnetic activity~\cite{clette2014}.
The recording of daily observations spans the years 1858 to 2018 and is provided by the World Data Center 
SILSO (Solar Influences Data Analysis Center) ~\cite{SILSO_Sunspot_Number}.
}

\section{Results}
\subsection{Model Systems}
\begin{figure}[htbp]
    \centering
    \includegraphics[width=\linewidth]{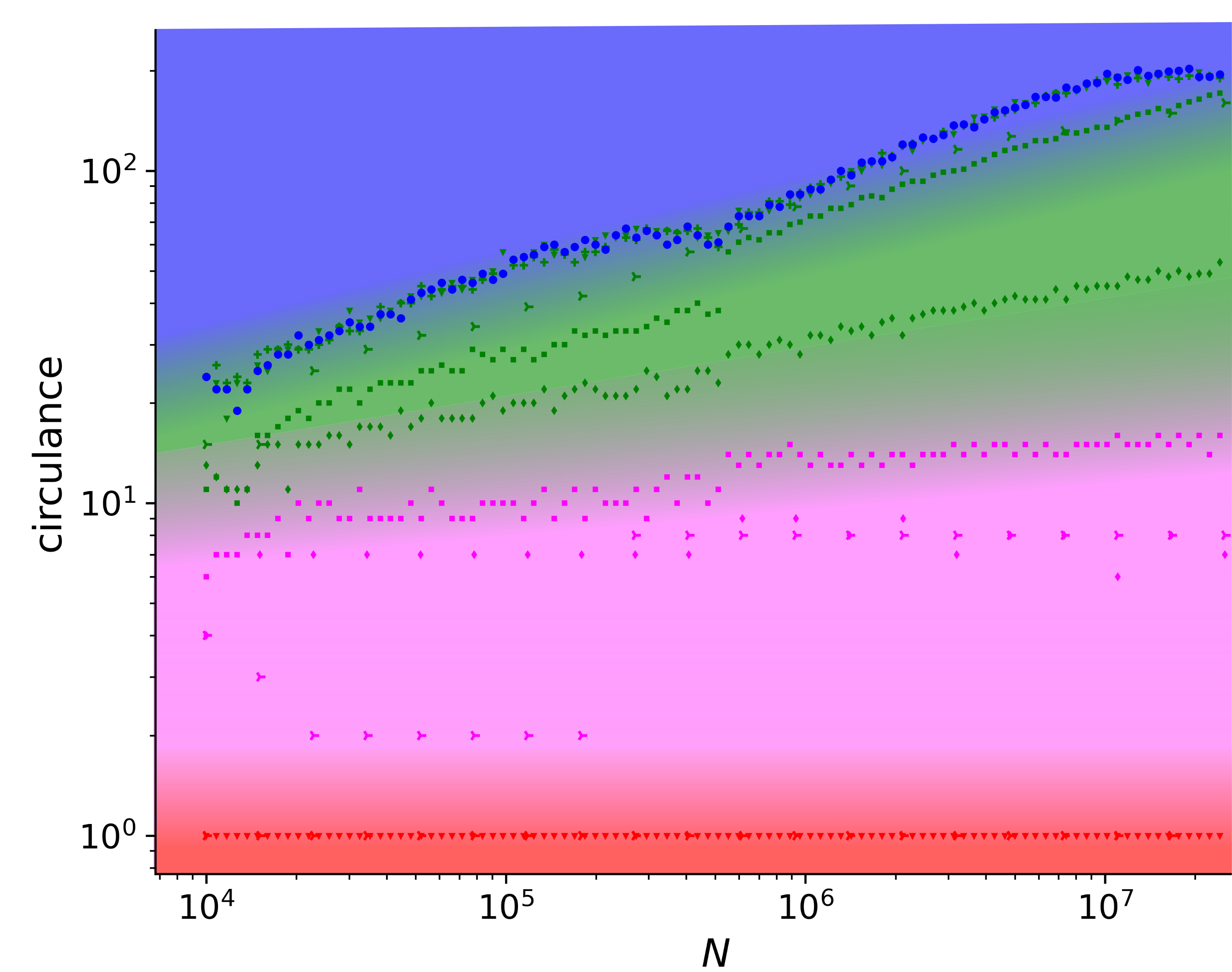}
    \caption{\com{Time series classification in the $\Qoppa-N$-plane with dynamical regimes ranging from periodic via quasiperiodic and chaotic to stochastic (marker color resp. red, green, pink, and blue). 
    Background colors schematically depict the continuum inferred empirically. 
    Overlapping regimes between chaos and stochasticity point to limited distinguishability of high-dimensional chaos and white noise.  
    Model systems are \com{associated with} different markers as follows: white noise - circle; generalized Hénon map - downward triangle; Zaslavskii map - plus; Hénon map - diamond; Lotka-Volterra system - square; Lorenz96 system -  rotated Y.
	Circulance values of time series from model systems with $N=10000$ from Table~\ref{tab:dynamics_circulance} accurately fit into the $\Qoppa-N$-plane.}}   
    \label{fig:schematic}
\end{figure}

We begin by demonstrating the merit of circulance by 
analyzing time series of length $N=10000$ from a range of canonical dynamical systems  
and summarize our findings in Table~\ref{tab:dynamics_circulance}. 
\com{The comparison against ground-truth and referencing against stochasticity 
allows a quantitative assessment of circulance's ability to correctly identify dynamical regimes.}
For strictly periodic time series,
circulance consistently~\cite{weird_henon} attains its minimal value ($\Qoppa=1$), reflecting the uniform circular structure of the corresponding OPTNs.
\com{For stochastic \com{time series},
we observe highest circulance values ($\Qoppa\geq18$), approaching the upper bound of the spectrum ($\Qoppa = d!$), which depends on the network size (number of vertices) and the length of the time series $N$ (see Eq.~\ref{eq:upper_bound}).} 
In comparison to periodic time series, quasiperiodic time series
yield slightly elevated circulance values ($2\leq \Qoppa\leq 4$), indicating modest deviations from perfect regularity.
In contrast, chaotic dynamics
exhibit intermediate circulance values ($4<\Qoppa<18$), capturing the increased diversity and complexity of ordinal pattern transitions. 
\com{Comparisons to further model systems (Appendix~\ref{app:mod}), accentuate that circulance takes on similar values for higher-dimensional chaos/hyperchaos (e.g., generalized Hénon map) and stochastic \com{time series} (Fig.~\ref{fig:schematic}).
This indicates a limitation of circulance -- in its current form -- due to it considering only one time series (from a specific system component or from an average over all components), assuming that this time series is representative of the system's dynamics.
}

Figure~\ref{fig:schematic} \com{also} demonstrates how the length of a time series $N$ impacts on the co-domain of circulance and thus on its ability to position time series along a continuous spectrum from regularity to randomness. 
\com{For shorter time series or lower embedding dimensions, the limited number of possible ordinal patterns restricts the capability of circulance to resolve dynamical regimes.} As the embedding dimension decreases, circulance of chaotic time series approaches that of stochastic \com{time series}, underscoring the importance of selecting the embedding dimension such that $N\sim d!^2$ to ensure a reliable \com{distinction} between dynamical regimes.\\
Since the calculation of $\qoppa_{\tau}$ scales approximately linear with the number of datapoints $N$ and an accurate \com{distinction} between dynamical regimes is already possible for a few thousand datapoints, the overall approach is rather computationally efficient~\cite{computationally_efficient}.

\begin{figure}[htbp]
        \centering
        \includegraphics[width=\linewidth]{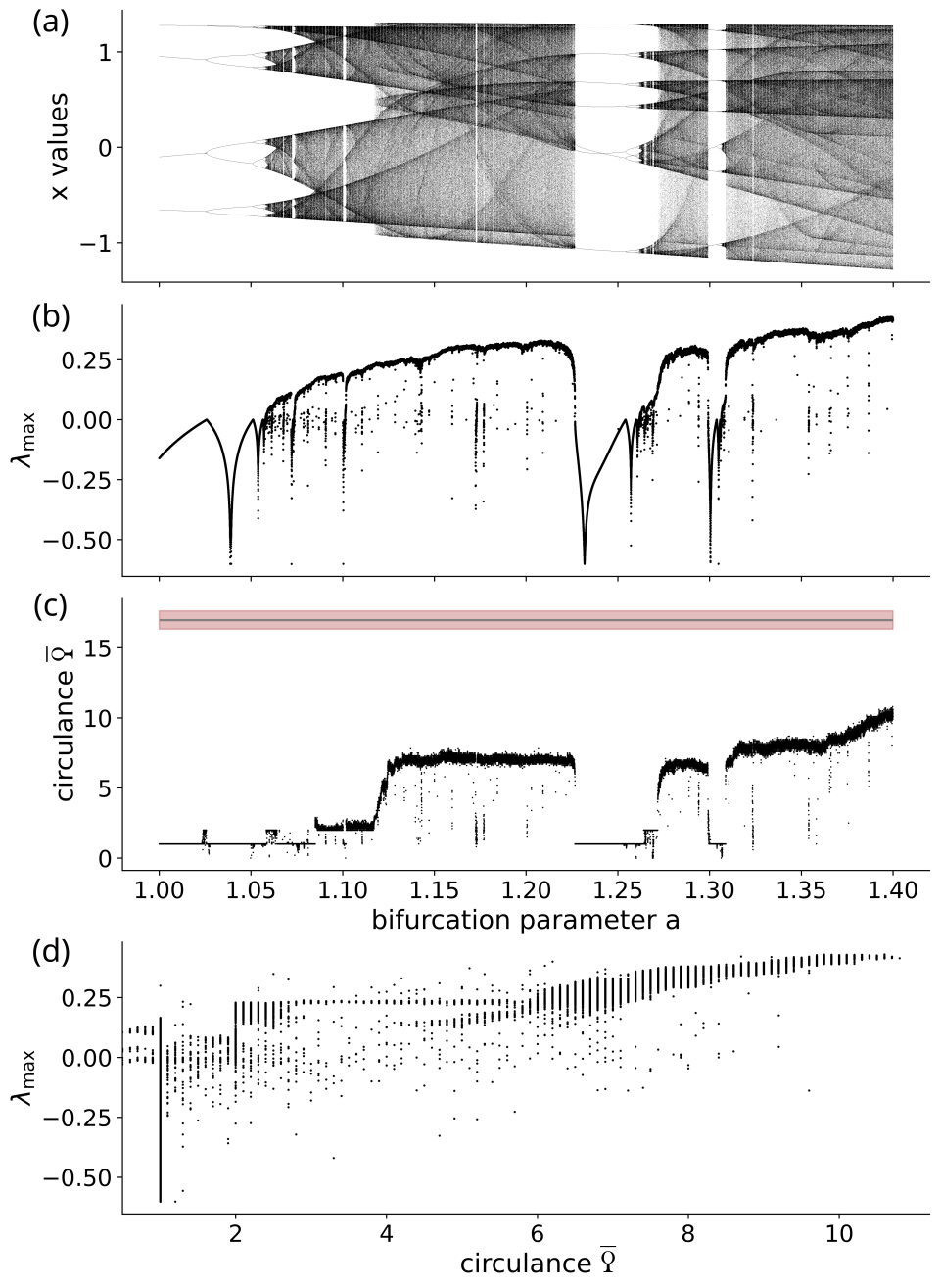}
        \caption{\com{Bifurcation diagram of the standard Hénon map (a).
        Mean largest Lyapunov exponent $\lambda_{\rm max}$~\cite{parlitz1992} (b) and \com{mean} circulance $\overline{\Qoppa}$ \, (c) of time series ($x$-component; $N=10000$) for different values of the bifurcation parameter $a$.
        Each datapoint is the mean of $10$ realizations of time series with the same bifurcation parameter $a$ but with different initial conditions. 
        The gray line and the red-shaded area in (c) indicate mean and standard deviation of circulance ($\overline{\Qoppa}=16.98\pm0.65$) of white noise time series of the same length and averaged over $1000$ realizations. 
        (d): Scatterplot of $\lambda_{\rm max}$ (from (b)) and $\overline{\Qoppa}$ (from (c)).}
        }
        \label{fig:henon_circ}
    \end{figure}

\com{We proceed with benchmarking circulance against the largest Lyapunov exponent that is often used to distinguish between periodic and chaotic dynamics.}
To do so, we \com{perform a detailed investigation of the standard Hénon map~\cite{Henon1976}, which is a relatively simple dynamical system, but which, upon iteration, produces extraordinarily complex phenomena.}
Figure~\ref{fig:henon_circ} illustrates the sensitivity of circulance to reflect transitions of the Hénon map as the bifurcation parameter $a$ is varied. 
\com{The alterations of circulance seen for a variation of the bifurcation parameter $a$ resemble the ones of the largest Lyapunov exponent, and both measures distinguish between periodic and chaotic regimes. 
In comparison to the largest Lyapunov exponent, the range of circulance values observed for chaotic regimes identified with the largest Lyapunov exponents (Fig.~\ref{fig:henon_circ} (c)) may allow to distinguish dynamics within these regimes in greater detail (e.g., distinguish quasiperiodicity from chaos).}

\subsection{Real World Systems}
\begin{figure*}[htbp]
	\centering
	\includegraphics[width=\linewidth]{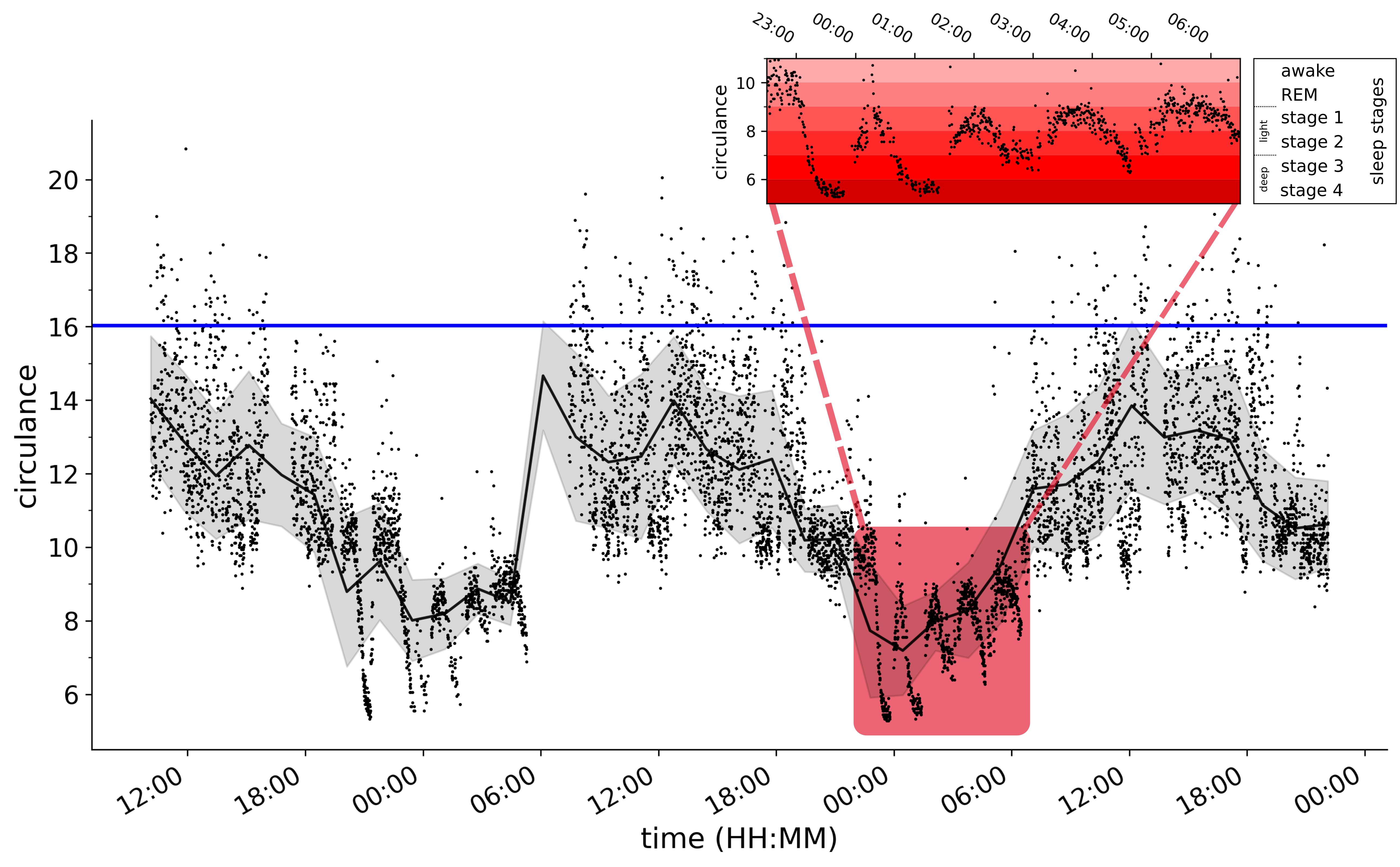}
	\caption{Time-dependent changes of circulance (mean over all electrodes) from a multi-day EEG recording. 
		The black line (grey shaded area) indicates the moving average (standard deviation) over 250 datapoints and serves as visual guidance.
		The inset schematically visualizes the potential association of ultradian rhythms during nighttimes with a sleep cycle and different sleep stages. 
		\com{The blue line indicates mean circulance of 100 stochastic time series with a length that equals the window size used for the sliding-window approach.}}
	\label{fig:circ_eeg}
\end{figure*}

\com{We proceed with an analysis of} multi-channel electroencephalographic (EEG) data from a healthy subject.
A comprehensive understanding of brain function requires careful characterization of its underlying, possibly nonlinear dynamics~\cite{andrzejak2001a}.
Investigations of EEG \com{time series} often render brain dynamics seemingly stochastic~\cite{pijn1991,prichard1994}, which may be due to the fact that the EEG only represents a one-dimensional projection of the brain's high-dimensional state-space.
Yet, EEG \com{time series} have also been observed to exhibit complex behavior characterized by nonlinear dynamics~\cite{stam2005,Tong2009}, and long-term EEG recordings are observed to also reflect influences by biological rhythms~\cite{Lehnertz2021a,Broehl2023a}, among others.
In this context, the notion of the brain as a system that exhibits chaotic dynamics remains controversially discussed by employing methods from nonlinear time series analysis in various studies in neuroscience~\cite{lehnertz2000,korn2003}. 
Using a sliding-window approach (window size: $20\unit[20]{s}$, $5120$ datapoints; no overlap), we computed circulance for each window -- within which the system is considered approximately stationary~\cite{lehnertz2017} --, thereby tracking temporal fluctuations in the dynamical regime of the brain's electrical activity.
These fluctuations exhibit a clear diurnal pattern (Fig.~\ref{fig:circ_eeg}):
circulance is elevated during daytime ($9\leq\Qoppa\leq24$), reflecting the brain's exposure to diverse sensory inputs and external perturbations during wakefulness.
Circulance decreases during nighttimes ($6\leq\Qoppa\leq12$), consistent with the emergence of a more regular, rhythmic activity during sleep.
Notably, 
we also observe the temporal evolution of circulance during nighttimes to exhibit pronounced oscillations with a period length of about $90$ minutes, which potentially reflects the characteristic cycling of sleep stages~\cite{aeschbach1993,carskadon2005}. 
Ranges of circulance values may correspond to different stages within the sleep cycle, such as transitions between slow-wave (deep) sleep and rapid eye movement (REM) sleep (inset in Fig.~\ref{fig:circ_eeg}).
The time-dependent changes of circulance thus provide a qualitative and quantitative view into the evolving (ir)regularity of brain activity, with potential utility for distinguishing behavioral and physiological states. 
While the interpretation of circulance in high-dimensional, complex systems such as the brain must be approached with caution -- given the limitations of single/multi-channel EEG/s as a proxy for the full system's dynamics -- the observed trends are consistent with prior empirical findings.
\com{Overall, our application to EEG data suggests the potential of circulance for uncovering \com{alterations of human brain dynamics.}}\\

\com{
Eventually, we demonstrate how circulance can possibly aid in improving understanding of solar magnetic activity, which requires careful characterization of its underlying, possibly nonlinear dynamics.
Investigations of sunspot number time series point to low-dimensional chaotic dynamics of the Sun's magnetic activity on shorter time scales and to seemingly stochastic dynamics on longer time scales~\cite{zhou2014,deng2015comparison}.
This may be due to the fact that sunspot observations only represent a low-dimensional projection of the complex high-dimensional magnetic field processes~\cite{letellier2006}.}
\begin{figure*}[htbp]
	\centering
	\includegraphics[width=\linewidth]{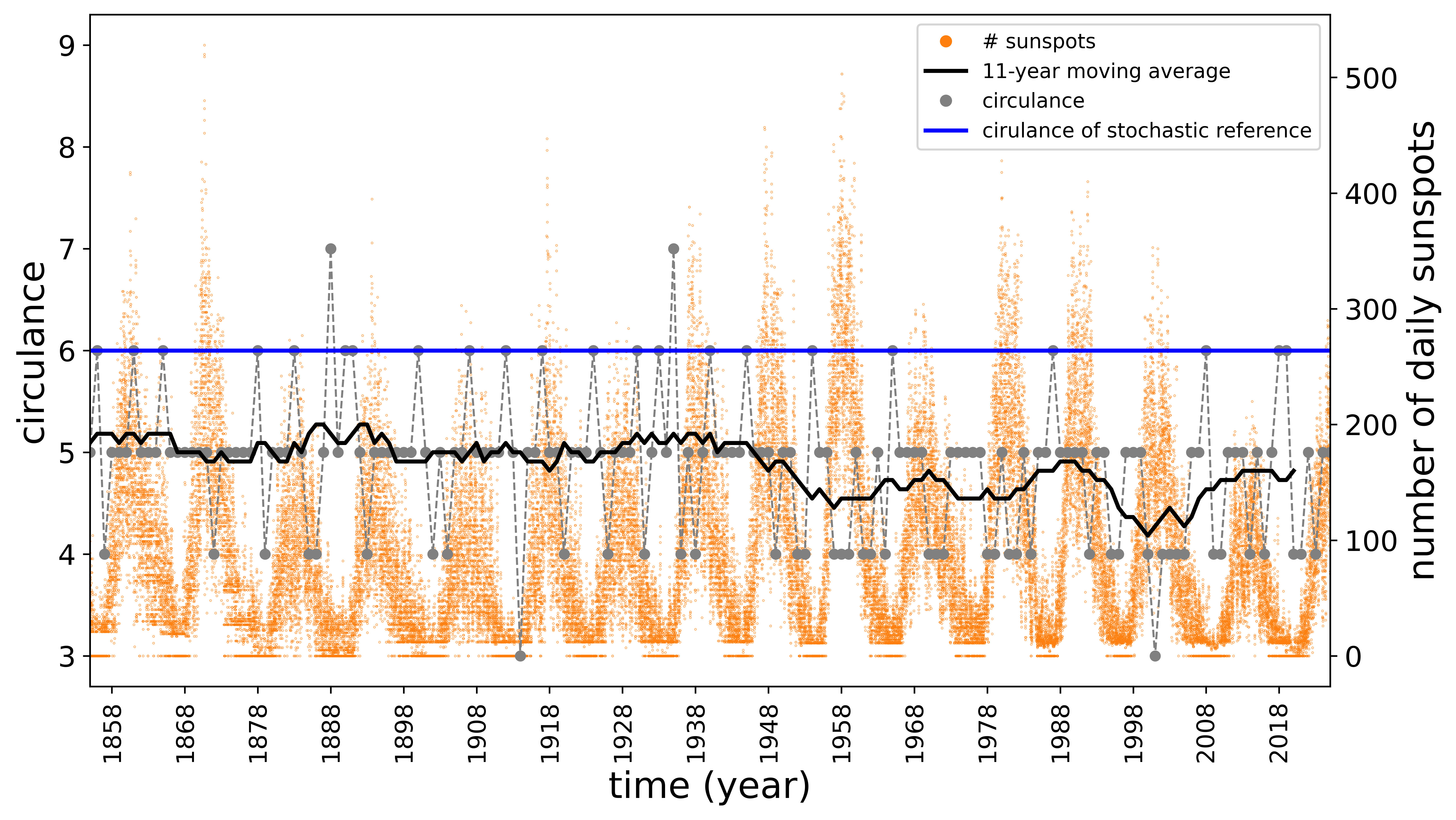}
	\caption{\com{Yearly changes of circulance (grey dots; dashed grey line for eye-guidance only) estimated from the daily number of sunspots (orange). 
	The black line represents the moving average over 11 years.}}
	\label{fig:circ_sunspot}
\end{figure*}  
\com{When calculated on yearly time periods employing a sliding-window approach (window size: 1 year, 365 datapoints; no overlap), circulance reveals systematic variations that do not necessarily correlate with solar cycle phases. 
Circulance takes values between $\Qoppa=3$ and $\Qoppa=7$, with a median value of $\Qoppa_\mathrm{med}=5$, which is smaller than the circulance value of a stochastic time series (white noise) of equal length ($\Qoppa=6$). 
This may indicate that inter-annual fluctuations resemble properties of deterministic chaos.
Notably, we observe the temporal evolution of circulance during individual $\sim11$-year sunspot cycles~\cite{hathaway2015} to exhibit pronounced oscillations with characteristic time scales of 2-4 years, which potentially reflects the underlying deterministic dynamics of the solar magnetic activity.
A local minimum of circulance ($\Qoppa=3$) can be observed in the middle (year 2001) of solar cycle 23, which started in August 1996 and ended in December 2008.
This is in accordance with prior research~\cite{deToma2004}, in which the cycle has been described as "magnetically simpler" than its closely preceding cycles (21 and 22) and therefore can be considered anomalous on a decadal timescale.
When the entire multi-decadal sunspot record is analyzed as a continuous time series without cycle-based segmentation, circulance attains a significantly higher value ($\Qoppa=39$) which is equal to the circulance value of a stochastic time series (white noise) of equivalent length.
This may suggest that the long-term evolution of solar magnetic activity, when viewed across multiple cycles, resembles a stochastic process rather than deterministic chaos. 
The scale-dependence may provide quantitative indications for the dual nature of solar magnetic variability: exhibiting deterministic characteristics at the inter-annual level yet displaying stochastic properties over extended multi-decadal periods~\cite{vashishtha2024}.
Temporal changes of circulance may thus provide both qualitative and quantitative insights into the evolving (ir)regularity of solar magnetic activity.
While the interpretation of circulance in high-dimensional complex systems such as the Sun must be approached with caution -- and further given the limitations of sunspot number records as proxies for the solar magnetic activity~\cite{clette2014} -- the observed indicative trends are consistent with prior empirical findings from nonlinear time series analyses of sunspot numbers~\cite{jevtic2001,letellier2006,zhou2014,vashishtha2024}.\\

Summarizing this section, the accurate characterization of predetermined dynamics of canonical model systems validates the capability of circulance for a time-series-based differentiation between periodic, quasiperiodic, chaotic and stochastic dynamics.
Our application to EEG and sunspot number time series suggests the potential of circulance for tracking alterations in the dynamics of complex systems that can be associated with known states and for placing their dynamics along a spectrum from regularity to randomness.

}

\section{Conclusion}
\com{We have introduced a scalar measure for classifying 
time series that is based on topological properties -- encoding the time series' regularity -- of directed ordinal pattern transition networks (OPTNs).
Circulance quantifies how much an OPTN deviates from a circular structure associated with strict periodicity.
This allows a differentiation} between strictly periodic, quasiperiodic, chaotic, and stochastic regimes, while providing a \com{quantitative} metric that positions time series along a spectrum from regularity to randomness.
Assuming that a time series adequately reflects the underlying system dynamics, circulance offers direct insight into the temporal organization and \com{evolution of complex dynamical systems.} 
Our \com{time-series-based} validation across a wide range of \com{predetermined dynamics of canonical model systems} demonstrate that circulance reliably identifies and distinguishes dynamical regimes.
\com{Our findings achieved with time-resolved analyses of circulance of real-world data hint to the possibility to} facilitate the detection of transient regimes, transitions \com{from quasiperiodicity} to chaos, or more detailed classifications of dynamical states.
Furthermore, integrating circulance into control strategies may support the stabilization or targeted alteration of desired dynamical states in complex systems.

Overall, circulance provides a powerful, interpretable, and computationally efficient means for positioning time series in a continuous spectrum from regularity to randomness.
\com{As a topology-based metric, it bridges nonlinear time series analysis and network theory, offering both theoretical insights and practical advantages for the study of complex systems \com{ranging from} physics \com{to} neuroscience. }
\com{Beyond nuanced classification, the OPTN framework and circulance open several promising avenues for future research.}
We anticipate that further developments and applications of circulance\com{, possibly even to other network representations of time series,} will advance our understanding of real-world dynamical systems.

\section*{Data Availability}
\com{The sunspot number dataset is publicly available from~\citet{SILSO_Sunspot_Number}.}
The EEG dataset presented in this article is not readily available because it contains information that could compromise the privacy of the research participant. 
Requests to access this dataset should be directed to the corresponding author.
\section*{Acknowledgements}
The authors would like to thank Christian Hechler and Manuel Lourenço for interesting discussions and for critical comments on earlier versions of the manuscript.
This work was supported by the Deutsche Forschungsgemeinschaft and the Verein zur Foerderung der Epilepsieforschung e. V. (Bonn).

\twocolumngrid
\noindent
\appendix
\counterwithin{figure}{section}

\section{Details of Model Systems}
\label{app:mod}

\vspace{0.3cm}

\com{
	If not stated otherwise, for the following systems we used the $x$($x_1$)-components as observables.
	
	\vspace{0.3cm}
	
	\noindent \textbf{Superimposed sine waves} (strictly periodic):
	\begin{equation*}
		x(w) = \sin(w) + a\sin(3w) + b\sin(13w),
	\end{equation*}
	with $w\in[\alpha,\beta)$, with $\alpha=-5$ and $\beta=5$ and parameter combinations: $(a,b) = (0,0)$, $(1,0)$, and $(1,1)$ \com{for the three cases shown in Table~\ref{tab:dynamics_circulance}}. Step size is $\frac{\beta-\alpha}{N}$ according to time series length $N$.
	
	\vspace{0.3cm}
	
	\noindent \textbf{Sawtooth} \com{time series} (strictly periodic):
	\begin{equation*}
		x(w) = w \bmod 1
	\end{equation*}
	with $w \in [\alpha, \beta)$ and $\alpha=0$ and $\beta=100$. Step size is $\frac{\beta-\alpha}{N}$  according to time series length $N$.
	
	\vspace{0.3cm}
	
	\noindent \textbf{Van der Pol oscillator} (periodic)\com{~\cite{van1920theory}}:
	\begin{align*}
		\dot{x} &= y,\\
		\dot{y} &= \mu(1-x^2)y - x,
	\end{align*}
	with damping parameters $\mu = 2$ \com{or} $\mu = 10$ \com{for the two cases shown in Table~\ref{tab:dynamics_circulance}}. Initial conditions: $(x(0), y(0)) = (0.1, 0.1)$.
}
\vspace{0.3cm}

\noindent \textbf{FitzHugh-Nagumo oscillator} (quasiperiodic)\com{~\cite{fitzhugh1961}}:
\begin{align*}
	\dot{x} &= x - \frac{x^3}{3} - y + I_{\mathrm{ext}},\\
	\epsilon\dot{y} &= x + a - by,\\
	I_{\mathrm{ext}} &= I_0 + A\sin(\omega t),
\end{align*}
with $a = 0.7$, $b = 0.8$, $I_0 = 0$, $A = 0.9$, $\omega = 0.02$, and time scale separation parameters $\epsilon = 1/12.5$ \com{or} $\epsilon = 1/2$ for the two cases shown in Table~\ref{tab:dynamics_circulance}. \com{Initial conditions ($x(0)$,$y(0)$) = ($-1.2,-0.6$).}

\vspace{0.3cm}

\noindent \textbf{Coupled FitzHugh-Nagumo oscillators} (quasiperiodic/chaotic)\com{~\cite{ansmann2013}}:
\com{
	\begin{align*}
		\dot{x}_i &= x_i(a - x_i)(x_i - 1) - y_i + \frac{\sigma}{O-1}\sum_{j=1}^{O} C_{ij}(x_j - x_i),\\
		\dot{y}_i &= b_i x_i - cy_i,
	\end{align*}
	with $O = 101$ oscillators, $i \in \{1, \ldots, O\}$, heterogeneous parameter $b_i = 0.006 + 0.008(i-1)/(O-1)$, and all-to-all coupling matrix $C_{ij} = 1$ for $i \neq j$, $C_{ii} = 0$. Initial conditions chosen uniformly from the interval $\left[0,100\right]$. Coupling strengths $\sigma \in \{0.002, 0.128, 0.628\}$, $a = -0.02651$, $c = 0.02$ for the three cases shown in Table~\ref{tab:dynamics_circulance}.}

\vspace{0.3cm}

\noindent \textbf{Hénon map} (quasiperiodic/chaotic)\com{~\cite{Henon1976}}:
\begin{align*}
	x(n+1) &= 1 + y(n) - ax(n)^2,\\
	y(n+1) &= bx(n),
\end{align*}
with $b = 0.3$ and bifurcation parameter $a \in \{1.1, 1.2, 1.4\}$ \com{for the three cases shown in Table~\ref{tab:dynamics_circulance} and $n=0,\dots,N-1$.} Initial conditions: $(x(0), y(0)) = (0, 0)$.

\vspace{0.3cm}
\noindent \textbf{Generalized Hénon map} (periodic/chaotic)\com{~\cite{baier1990maximum}}:

\begin{align*}
	\begin{pmatrix} x_1(n+1) \\ x_i(n+1)  \end{pmatrix} = \begin{pmatrix} a-x_{k-1}(n)^2-bx_k(n) \\ x_{i-1}(n) \end{pmatrix},
\end{align*}
with $i\in\{2,3,\dots,k\}$, $k=50$, $b = 0.1$ and $a\in\{1.25,1.75\}$ \com{for the two cases shown in Fig.~\ref{fig:schematic}}. \com{ $n=0,\dots,N-1$. Initial conditions chosen uniformly from $\left[0,1\right]$, respectively.}

\vspace{0.3cm}

\noindent \textbf{Lotka-Volterra model} (quasiperiodic/chaotic)\com{~\cite{vano2006chaos}}:
\begin{align*}
	\dot{x}_i = r_ix_i\left(1-\sum_{j=1}^NQ_{ij}x_j\right) 
\end{align*}
\com{with $i\in\{0,1,2,3\}$ and
	
	\begin{align*}
		r = \begin{pmatrix} 1.0 \\ 0.72 \\ 1.53\\1.27  \end{pmatrix} \hspace{1mm}\mathrm{and}\hspace{1mm} Q = \begin{pmatrix} 1.0 &1.09s&1.52s&0.0 \\ 0.0&1.0&0.44s&1.36s \\ 2.33s&0.0&1.0&0.47s\\1.21s&0.51s&0.35s&1.0  \end{pmatrix},
	\end{align*}
	with $s=0.97$ or $s=1.0$ respectively for the two cases shown in Fig.~\ref{fig:schematic}.}
\com{Initial conditions chosen uniformly from $\left[0.3,0.7\right]$, respectively.}

\vspace{0.3cm}

\com{\noindent \textbf{Zaslavskii map} (chaotic)\com{~\cite{zaslavsky1978simplest}}:
\begin{align*}
	x(n+1) &= (x(n)+\nu(1+\mu y(n))\\
	&+\epsilon\nu\mu\cos(2\pi x(n)))\hspace{2mm}\rm mod\hspace{1mm} 1\\
	y(n+1) &= e^{-r}(y(n)+\epsilon\cos(2\pi x(n)))
\end{align*}
with $\nu = 0.095$, $\mu = 20$, $\epsilon = 0.4$ and $r=1$ \com{and $n=0,\dots,N-1$. Initial conditions: $(x(0), y(0)) = (0, 0)$.}}
\vspace{0.3cm}

\noindent \textbf{Lorenz oscillator} (quasiperiodic)\com{~\cite{lorenz1963}}:
\begin{align*}
	\dot{x} &= \sigma(y - x)\\
	\dot{y} &= x(\rho - z) - y\\
	\dot{z} &= xy - \beta z
\end{align*}
with $\sigma = 10$, $\beta = 8/3$, and control parameter $\rho \in \{155, 80\}$ for the two cases shown in Table~\ref{tab:dynamics_circulance}. Initial conditions: $(x(0), y(0), z(0)) = (1, 1, 1)$.

\vspace{0.3cm}
\noindent \textbf{Lorenz 96 model} (periodic/quasiperiodic/chaotic)\com{~\cite{lorenz1996predictability}}:
\begin{equation*}
	\dot{x}_i = (x_{i+1} - x_{i-2})x_{i-1} - x_i + F,
\end{equation*}
\com{
	with $i \in \{1, 2, 3, 4, 5\}$ and periodic boundary conditions:\com{
		\[
		x_{i-5} = x_i = x_{i+5}
		\]
	} where $F\in\{3,5,8\}$ is a constant forcing term (commonly $F=3$ for periodic, $F=5$ for quasiperiodic and $F=8$ for chaotic behavior; cf. Fig.~\ref{fig:schematic}). Initial conditions are at equilibrium ($x_i(0)=1$).
	
	\vspace{0.3cm}
	
	\noindent \textbf{White noise} (stochastic):
	Gaussian white noise $x(t) \sim \mathcal{N}(0, 1)$ generated using NumPy's \texttt{numpy.random.normal} function.}

\vspace{0.3cm}

\com{
\section{Pseudocode}

The derivation of OPTNs from time series and the computation of circulance is openly accessible as Python code on GitHub~\cite{opytn2025}.
In the following, we additionally provide pseudocode for easy and quick comprehension.

\subsection{Derivation of Ordinal Patterns}
\FloatBarrier 
\begin{figure}[htbp]
\begin{algorithm}[H]
	\caption{Derive Ordinal Patterns from Time Series}
	\begin{algorithmic}[1]
		\Require Time series $TS \in \mathbb{R}^n$, embedding dimension $d \in \mathbb{N}$, embedding delay $\tau \in \mathbb{N}$
		\Ensure Set of ordinal patterns $\mathcal{P}$
		\Procedure{ordinal\_patterns}{$TS$}
		\State $TS \leftarrow \text{round}(TS, 7)$ \Comment{Round to 7 decimal places}
		\State $n \leftarrow |TS|$
		\State $\mathcal{P} \leftarrow \emptyset$
		\For{$i = 0$ \textbf{to} $n - (d-1)\tau - 1$}
		\State $w \leftarrow (TS_i, ts_{i+\tau}, TS_{i+2\tau}, \ldots, TS_{i+(d-1)\tau})$ \Comment{Extract window}
		\State $\pi \leftarrow \text{argsort}(w)$ \Comment{Get permutation indices}
		\State $\mathcal{P} \leftarrow \mathcal{P} \cup \{\pi\}$
		\EndFor
		\State \Return $\mathcal{P}$
		\EndProcedure
	\end{algorithmic}
\end{algorithm}
\end{figure}
\FloatBarrier

\subsection{Determine Embedding Dimension}
\FloatBarrier 
\begin{figure}[H]
\begin{algorithm}[H]
	\caption{Determine Embedding Dimension}
	\begin{algorithmic}[1]
		\Require Time series length $N \in \mathbb{N}$
		\Ensure Embedding dimension $d \in \mathbb{N}$
		\Procedure{give\_d}{$N$}
		\State $n \leftarrow 1$, $i \leftarrow 1$
		\While{$n < \sqrt{N}$}
		\State $i \leftarrow i + 1$
		\State $n \leftarrow n \times i$
		\EndWhile
		\State \Return $i$
		\EndProcedure
	\end{algorithmic}
\end{algorithm}
\end{figure}
\FloatBarrier

\subsection{Determine Maximum Embedding Delay}
\FloatBarrier 
\begin{figure}[H]
\begin{algorithm}[H]
	\caption{Determine Maximum Embedding Delay}
	\begin{algorithmic}[1]
		\Require Time series $TS \in \mathbb{R}^n$
		\Ensure Maximum embedding delay $\tau_{\max} \in \mathbb{N}$
		\Procedure{find\_maxtau}{$TS$}
		\State $n \leftarrow |TS|$
		\State $\tau \leftarrow 2$
		\State $d \leftarrow \text{give\_d}(n)$ 
		\While{$(d-1)\tau < 0.1n$} \Comment{Ensure at most 10\% loss of data}
		\State $\tau \leftarrow \tau + 1$
		\EndWhile
		\State \Return $\tau - 1$
		\EndProcedure
	\end{algorithmic}
\end{algorithm}
\end{figure}
\FloatBarrier

\subsection{Construction of Ordinal Pattern Transition Networks}
\FloatBarrier 
\begin{figure}[H]
\begin{algorithm}[H]
	\caption{Create Adjacency Matrix of OPTN from Ordinal Patterns}
	\begin{algorithmic}[1]
		\Require Time series $TS \in \mathbb{R}^n$, embedding dimension $d \in \mathbb{N}$, embedding delay $\tau \in \mathbb{N}$
		\Ensure adjacency matrix $A \in \{0,1\}^{d \times d}$
		\Procedure{OPTN}{$TS$}
		\State $\mathcal{S} \leftarrow \text{ordinal\_patterns}(TS, d, \tau)$ \Comment{Extract ordinal patterns}
		\State $\mathcal{U}, \mathcal{I} \leftarrow \text{unique\_with\_indices}(\mathcal{S})$ \Comment{Get unique patterns and indices}
		\State $m \leftarrow |\mathcal{U}|$
		\State Initialize adjacency matrix $A$ with zeros
		\For{$j = 0$ \textbf{to} $|\mathcal{I}| - 2$}
		\State $i_{\text{from}} \leftarrow \mathcal{I}[j]$
		\State $i_{\text{to}} \leftarrow \mathcal{I}[j+1]$
		\If{$i_{\text{from}} \neq i_{\text{to}}$}
		\State $A[i_{\text{from}}, i_{\text{to}}] \leftarrow 1$\Comment{replace 0 with 1 if transition occurs}
		\EndIf
		\EndFor
		\State \Return $A$
		\EndProcedure
	\end{algorithmic}
\end{algorithm}
\end{figure}
\FloatBarrier

\subsection{Main Circulance Algorithm}
\FloatBarrier 
\begin{figure}[H]
\begin{algorithm}[H]
	\caption{Computation of Circulance}
	\begin{algorithmic}[1]
		\Require Time series $TS \in \mathbb{R}^n$, embedding dimension $m$ (optional), possible embedding delays $\Delta$ (optional)
		\Ensure Circulance value $c \in \mathbb{N}$, critical delay $\tau_c \in \mathbb{N}$, embedding dimension $d \in \mathbb{N}$
		\Procedure{Circulance}{$x$}
		
		\State $d \leftarrow \text{give\_d}(|TS|)$
		
		\State $\Delta \leftarrow \{2, 3, \ldots, \text{find\_maxtau}(TS)\}$

		\State $Q \leftarrow \emptyset$ \Comment{Initialize circulance values list}
		
		\For{each $\tau \in \Delta$}
		\State $A \leftarrow \text{OPTN}(TS, d, \tau)$ \Comment{Get adjacency matrix}
		
		\State $\text{outdeg} \leftarrow \sum_{j=1}^{|A|} A_{i,j}$ for all $i$ \Comment{Calculate out-degrees}
		\State $\text{indeg} \leftarrow \sum_{i=1}^{|A|} A_{i,j}$ for all $j$ \Comment{Calculate in-degrees}
		
		\State $q \leftarrow |\{\text{outdeg}\} \cap \{\text{indeg}\}|$ \Comment{Intersection of degree sets}
		\State $Q \leftarrow Q \cup \{q\}$
		\EndFor
		
		\State $i^* \leftarrow \arg\min_{i} Q[i]$ \Comment{Find index of minimum circulance}
		\State $c \leftarrow Q[i^*]$ \Comment{Minimum circulance value}
		\State $\tau_c \leftarrow \Delta[i^*]$ \Comment{Critical embedding delay}
		
		\State \Return $c, \tau_c, d$
		\EndProcedure
	\end{algorithmic}
\end{algorithm}
\end{figure}\hfill
}
\FloatBarrier

\begin{thebibliography}{84}
\bibitem [{wei()}]{weird_henon}%
	\BibitemOpen
	\href@noop {} {}\bibinfo {note} {In some distinct and rare cases within
		periodic regimes of the Hénon map, the respective cirulance value can be
		zero.}\BibitemShut {Stop}%
\bibitem [{com()}]{computationally_efficient}%
	\BibitemOpen
	\href@noop {} {}\bibinfo {note} {Estimating cirulance for a time series of
		$10^4$ data points can be performed in few milliseconds on a PC with a CPU
		clock rate of 3.7 GHz.}\BibitemShut {Stop}%
\end{thebibliography}
%

\end{document}